\documentclass[pre,superscriptaddress,reprint,showpacs,onecolumn,notitlepage,nobalancelastpage, eqsecnum]{revtex4-1}

\usepackage{graphicx} 
\usepackage{tikz}
\usepackage{amsmath}
\usepackage{amssymb}
\usepackage{color}
\usepackage{float}
\usepackage{epstopdf}
\usepackage[normalem]{ulem}
\usepackage{pgfplots,pgfplotstable}
\usepackage{soul}
\usepackage{accents}
\usepackage{nameref}
\usepackage{balance}
\usepackage{upgreek}
\usepackage[all]{nowidow}
\usepackage{comment}
\usepackage{dsfont}
\usepackage{cleveref}
\usepackage{array} 				
\usepackage{booktabs} 			
\usepackage{ctable}
\newcolumntype{q}[1]{>{\setlength{\parindent}{0em}}p{#1}}

\crefformat{section}{\S#2#1#3} 
\crefformat{subsection}{\S#2#1#3}
\crefformat{subsubsection}{\S#2#1#3}

\makeatletter
\newcommand*{\balancecolsandclearpage}{%
  \close@column@grid
  \clearpage
  \twocolumngrid
}
\makeatother

\usetikzlibrary[shadings]
\usetikzlibrary{shapes}
\usetikzlibrary{snakes}
\usepackage{bm}
\usepgfplotslibrary{fillbetween}
\tikzset{
    axis break gap/.initial=10mm
}
\usetikzlibrary{patterns}

\pgfplotsset{compat=newest}

\sethlcolor{white}

\newcounter{marknumber}
\pgfplotsset{
    error bars/every nth mark/.style={
        /pgfplots/error bars/draw error bar/.prefix code={
            \pgfmathtruncatemacro\marknumbercheck{mod(floor(\themarknumber/2),#1)}
            \ifnum\marknumbercheck=0
            \else
                \begin{scope}[opacity=0]
            \fi
        },
        /pgfplots/error bars/draw error bar/.append code={
            \ifnum\marknumbercheck=0
            \else
                \end{scope}
            \fi
            \stepcounter{marknumber}    
        }
    }
}

\AtBeginDocument{
\heavyrulewidth=.08em
\lightrulewidth=.05em
\cmidrulewidth=.03em
\belowrulesep=.65ex
\belowbottomsep=0pt
\aboverulesep=.4ex
\abovetopsep=0pt
\cmidrulesep=\doublerulesep
\cmidrulekern=.5em
\defaultaddspace=.5em
}

\newcommand{\vett}[1]{\mathbf{#1}}

\newcommand {\tr} {\mbox{\rm tr\,}}

\newcommand {\dd}{\mathrm{d}}
\newcommand{\cg}{\textnormal{\textsl{g}}}
{\left\lbrace\begin{array}{@{}l@{}}}%
{\end{array}\right.}

\newcommand{\dph}[1]{\textcolor{blue}{\emph{[#1]}}}

\newenvironment{myindentpar}[2]%
{\begin{list}{}%
         {\setlength{\leftmargin}{#1}\setlength{\rightmargin}{#2}}%
         \item[]%
}
{\end{list}}

\begin{document}

\title{
The nonlinear buckling behavior of a complete spherical shell under uniform external pressure and homogenous natural curvature}

\author{Douglas P. Holmes}
\email{dpholmes@bu.edu}
\author{Jeong-Ho Lee}
\author{Harold S. Park}
\affiliation{
Department of Mechanical Engineering, Boston University, Boston, MA, 02215.
}%

\author{Matteo Pezzulla}
\email{matteo.pezzulla@epfl.ch}
\affiliation{Institute of Mechanical Engineering, \'Ecole Polytechnique F\'ed\'erale de Lausanne, Lausanne CH-1015
}%

\date{\today}

\begin{abstract}
In this work, we consider the stability of a spherical shell under combined loading from a uniform external pressure and a homogenous natural curvature. Non--mechanical stimuli, such as one that tends to modify the rest curvature of an elastic body, are prevalent in a wide range of natural and engineered systems, and may occur due to thermal expansion, changes in pH, differential swelling, and differential growth. Here, we investigate how the presence of both an evolving natural curvature and an external pressure modifies the stability of a complete spherical shell. We show that due to a mechanical analogy between pressure and curvature, positive natural curvatures can severely destabilize a thin shell, while negative natural curvatures can strengthen the shell against buckling, providing the possibility to design shells that buckle at or above the theoretical limit for pressure alone, {\em i.e.} a {\em knock--up factor}. These results extend directly from the classical analysis of the stability of shells under pressure, and highlight the important role that non--mechanical stimuli can have on modifying the the membrane state of stress in a thin shell. 
\end{abstract}

\maketitle


\section{Introduction}


One of the great engineering challenges of the 18$^{th}$ century was the accurate determination of a ship's longitudinal location during long maritime voyages. At the heart of this challenge was the lack of a clock that could keep time accurate time while remaining unaffected by variations in temperature, pressure, and humidity. This problem was eventually solved by John Harrison, a clockmaker, who compensated for temperature changes with his {\em Thermometer Kirb}, and invention well known to today's engineers as the bimetallic strip. The now classical understanding of how a non--mechanical stimulus, such as temperature, imparts curvature of a bimetallic strip was not understood mechanically for another 160 years following Harrison's invention, when the bimetal was analyzed by Timoshenko~\cite{Timoshenko1925}. Timoshenko's result is now familiar to all mechanical engineers: under a homogenous temperature change, the two metals expand by different amounts, and the bimetal bends to adopts a {\em natural curvature} with residual thermal stresses in 3D~\cite{Ozakin2010}. It is now well understood that environmental conditions, such as changes in temperature, pH, and humidity may induce a non--mechanical stimulus in most engineering materials, and these effects can significantly alter the shape of a structure. For example, elastomers and gels, crosslinked either chemically or through entanglement, will swell in a favorable solvent, and this swelling can cause a significant increase in their volume -- sometimes surpassing several hundred percent~\cite{Hong2008, Lucantonio2013}. The differential swelling of a bilayer gel provides an extension of Timoshenko's analysis of a heated bimetallic strip to include nonlinearities that emerge from the large stretching strains~\cite{Lucantonio2014a}. Although the physics behind swelling and thermal expansion are quite different from each other, the similarities in these phenomena highlight the underlying geometric connection of non--mechanical stimuli to local volume changes within a material.

The effect that residual stress has on thin elastic structures also has a long history. Stoney analyzed the stress in the deposition of thin metallic films~\cite{Stoney1909, Freund1999}, a phenomenon that will lead to the bowing of the silicon wafers these films may be deposited on. Perhaps unsurprisingly, these affects amount to an extension of Timoshenko's bimetal analysis from beams to plates, but of course the geometric differences between a beam and a plate are not insignificant. Homogenous heating of a bimetal plate will endow the plate with a homogenous natural curvature, causing it to bend into the segment of a spherical cap, and thus adopting a positive Gaussian curvature. This change in Gauss curvature comes at the cost of stretching the plate's middle surface. Eventually, the energetic cost for the plate to bend into a cylinder becomes lower than the cost to continue bending into a spherical cap, and so the bowing wafer will buckle into a cylindrical shape~\cite{Mansfield1962, Mansfield1965,Masters1993, Salamon1995, Freund2000, Seffen2007}. This phenomenon is perhaps familiar to those who have cooked in the oven with a metallic baking sheet, as it may buckle and warp when heated above a certain temperature. Similar warping plagued the curing of laminated, fiber--reinforced composites~\cite{Hyer1981a,Hyer1981b, Hyer1982, Hamamoto1987}. Recently, researchers have targeted this bilayer buckling instability as a means for creating morphable, shape--shifting structures~\cite{Seffen2011, Chen2012, Seffen2013, Pezzulla2016, Jiang2018}. In general, these pre--stressed bodies have no {\em stress--free configuration}, and modeling the static and dynamic shapes of such structures led to an alternate formulation of elasticity theory built on a different definition of the strain tensor -- one that did not measure strain from the original configuration of the elastic body, but rather from a configuration that would render the elastic body stress--free. Termed {\em incompatible} or {\em non--Euclidean elasticity}, theoretical development began in the mid--twentieth century to describe incompatibility resulting from material defects, such as disclinations, dislocations, and point--defects~\cite{Nye1953, Kondo1955, Bilby1955, Wang1968, Kroner1981}.  Building on ideas from plasticity theory, where modeling of incompatible elasticity began with a multiplicative decomposition of the deformation gradient~\cite{Kondaurov1987, Takamizawa1987}, classical shell mechanics was extended recently to model bodies that do not possess a stress-free configuration~\cite{Gurtin2010, AmarGoriely2005,Goriely2005}, leading to the so--called non--Euclidean plate~\cite{Efrati2009} and shell models~\cite{Efrati2010, Pezzulla2017}. This approach to describe bodies that remain stressed even in the absence of externally applied mechanical forces has been used to study growth~\cite{Goriely2005, Yavari2010}, thermal expansion~\cite{Ozakin2010}, humidity--induced expansion and drying~\cite{Armon2011}, and gel swelling~\cite{Klein2007, Efrati2007, Efrati2009b, Efrati2011, Gemmer2013, Pezzulla2015, Pezzulla2016, Pezzulla2018}.


In looking beyond engineered materials, it quickly becomes apparent that Nature uses non--mechanical and internal stimuli to locally and globally change the curvature of thin and soft materials in a variety of ways -- a process that generates {\em natural} or {\em spontaneous curvature}. Lipids induce a global spontaneous curvature in the formation of vesicles~\cite{Helfrich1973, Helfrich1974, Miao1994, Capovilla2003, Steigmann2003}, they couple with the cytoskeleton of red blood cells to influence their shape~\cite{HW2002}, and they interact with proteins to locally cause curvature changes in the cellular membrane~\cite{Zimmerberg2006}. Differential growth rates in plant stems, branches, and roots induce curvature changes that provide the main mechanism responsible for their {\em gravitropic response}, {\em i.e.} their ability to redirect their growth direction vertically~\cite{Hejnowicz1997, Galland2002, Moulia2009, Bastien2013, Bastien2014, Goriely2017}. The coupling of local curvature and local stress may be necessary to enable the stable, elongating growth of cylindrical shells, such as {\em E. coli}~\cite{Mosleh2018}. Internal curvature changes in the leaves of the Venus flytrap trigger their rapid closure, enabling the plant to capture prey~\cite{Forterre2005}. Local curvature changes enable the ventral furrow formation in {\em Drosophila} embryos (small fruit flies)~\cite{Heer2017} and the eversion of {\em Volvox} embryos~\cite{Hohn2015, Haas2015} -- drastic morphological changes that are essential to their morphogenesis. Such curvature changes can be actively triggered, as is seen with the blooming of a lily~\cite{Liang2011}, and in the active stresses that underlie the gyrification of the cerebral cortex~\cite{Tallinen2014}. Many of these naturally occurring structures are both thin and curved, raising the question of how an evolving natural curvature effects the stability of thin shells. Indeed, the coupling of membrane stresses with curvature or twisting stresses was recently found to alter the fracture properties of thin rods, such as in the fracture cascades of dry spaghetti~\cite{Heisser2018}, calling further attention to the question of how these stresses may affect the instabilities of shells.  

The stability of thin shells under mechanical loads such as pressure and compression was an incredibly active area of research throughout the 20$^{th}$ century, in part because of an apparent disconnect between theoretical predictions and experimental evidence that arose from the extreme imperfection sensitivity of thin shells~\cite{Hutchinson1967, Elishakoff2014}. No attempt to survey the vast literature that lay in the wake of this controversy is made here, other than to note the excellent recent overview by Hutchinson~\cite{Hutchinson2016}. This review begins by recalling Koiter's contributions to the field in developing a nonlinear shell theory that will be used extensively in this paper~\cite{Sanders1963, Koiter1966,Koiter1967, Budiansky1968}. Recent work on the pressure buckling of shells has aimed to connect the geometric role of imperfections on the critical buckling pressure~\cite{Lee2016b}, and ways to probe these imperfections and energy barriers to shell buckling~\cite{Thompson2016, Hutchinson2017, Thompson2017, Marthelot2017, Virot2017}.  Pezzulla {\em et al.} first drew an analogy between pressure and natural curvature when considering the affects of natural curvature on open and closed elastic shells~\cite{Pezzulla2018}. This analogy enables one to consider this internal stimulus in a mechanical way by means of a {\em curvature potential}, which captures the deformation of a shell in it fundamental state under an evolving natural curvature.  Stability, of course, must not be analyzed by the character of this potential, but rather by the character of the stress state of deformed shell~\cite{Budiansky2013}. That work showed that an evolving natural curvature in the absence of pressure affects both the membrane and bending states of stress, the latter of which is typically neglected when considering a spherical shell loading by a uniform external pressure~\cite{koiter1969}. In contrast, in the absence of an evolving natural curvature, the buckling of a spherical shell is characterized by the familiar, classical result from Zoelly for the critical buckling pressure as~\cite{Zoelly1915}
\begin{equation}
p_c = \frac{2E}{\sqrt{3(1-\nu^2)}}\left(\frac{h}{R}\right)^2,
\end{equation}
where $E$ is Young's elastic modulus, $\nu$ is Poisson's ratio, $h$ is shell thickness, and $R$ is the radius of curvature of the shell. As we will show in this paper, for a spherical shell under both a uniform external pressure and a stimulus that induces a homogenous natural curvature, the buckling pressure is modified such that
\begin{equation}
p_\kappa=2 E\left(\frac{h}{R}\right)^2 \left[\sqrt{\frac{1}{3 \left(1-\nu ^2\right)}-\frac{\kappa^2 h^2 (1+\nu)^2}{36 \left(1-\nu ^2\right)}}-\frac{\kappa h(1+2 \nu)}{12 (1-\nu )}\right],
\end{equation}
where $\kappa$ is the magnitude of natural curvature, a quantity that can be either positive or negative. It is immediately clear that in the absence of a natural curvature stimulus, {\em i.e.} $\kappa=0$, we recover the classical buckling pressure of a spherical shell obtained by Zoelly. As we will see, the consequences of a shell under combined pressure and curvature loading are compelling -- natural curvature can act to destabilize a shell, or strengthen it against buckling providing the possibility to design pressure vessels with a {\em knock--up} factor. Since natural curvature can be imparted by differential swelling or through the heating of bimetal shells, it may become an intriguing design parameter in the processing and characterization of thin elastic shells. 

We outline the paper as follows: in \S\ref{stability} we review the stability criteria for elastic bodies under conservative loading. In \S\ref{shelltheory}, we derive Koiter's equations for the strain energy of a spherical shell. In \S\ref{fundamental}, we identify the total potential energy of a spherical shell in the fundamental state when loaded by a combination of uniform pressure and homogenous natural curvature, taking care to retain the contribution of the natural curvature to the membrane and bending prestress. In \S\ref{reduction}, we reduce the total potential energy by decomposing the tangential displacement field of the shell, focusing our efforts on the additional contributions from the presence of a nonzero natural curvature. Finally, in \S\ref{stabilityAnalysis}, we expand the total potential energy in spherical harmonics, perform linear stability analysis, and arrive at the critical buckling pressure for shell's under combined pressure and curvature loading. We compare these analytical results to numerical results from a 1D axisymmetric shell model (\ref{1D}), as well as a 2D model that allows for nonaxisymmetric deformations (\ref{2D}).

\section{Stability Criteria}
\label{stability}
In this section, we will outline the stability criteria for elastic bodies under conservative loading. We will closely follow the mathematical treatment by Koiter~\cite{koiter}, however if a more conceptual understanding of the fundamental theorems of elastic stability are desired, we recommend the works by Hunt and Thompson\cite{thompson, thompsonRed}. Consider a three dimensional body $\mathcal{B}=\bm{R}(x^1, x^2, x^3)$ embedded in $\mathds{R}^3$. We adopt the standard notation, letting Latin indices $i,j,\ldots \in (1,2,3)$ and Greek indices $\alpha, \beta,\ldots \in (1,2)$. The potential energy functional of an elastic body $\mathcal{V}$ is composed of the elastic potential per unit volume $W(\bm{\gamma})$ integrated over the material volume plus the potential of the external loads $P[\vett{\Psi(\vett{x})}]$, 
\begin{equation}
\mathcal{V}[\vett{\Psi(\vett{x})}]=\int_V W(\bm{\gamma}) \ \dd V + P[\vett{\Psi(\vett{x})}].
\end{equation}
To better understand the problem of stability, we will consider each term in this energy functional more carefully by expanding them in a Taylor series about their value in the fundamental, deformed state I. Beginning with the elastic potential per unit volume, we find
\begin{equation}
\label{eqW}
W(\bm{\gamma})=\left(\frac{\partial W}{\partial \gamma_{ij}}\right)_I\gamma_{ij}+\frac{1}{2}\left(\frac{\partial^2 W}{\partial \gamma_{ij} \partial \gamma_{kl}}\right)_I\gamma_{ij}\gamma_{kl}+\ldots,
\end{equation}
where the first term can be further examined by virtue of its first variation as
\begin{equation}
\delta W=\frac{1}{2}\left(\frac{\partial W}{\partial \gamma_{ij}}+\frac{\partial W}{\partial \gamma_{ji}}\right)_I \delta \gamma_{ij}\equiv S^{ij}\delta\gamma_{ij}.
\end{equation}
Here we have introduced a symmetric stress tensor $S^{ij}=S^{ji}$, {\em i.e.} the second Piola--Kirchhoff stress tensor, and the Green--Lagrange strain tensor $\gamma_{ij}$, which is given in terms of the displacement vector $\vett{\Psi}(\vett{x})$ as
\begin{equation}
\gamma_{ij}=\frac{1}{2}(\Psi_{i,j}+\Psi_{j,i}+\Psi_{h,i}\Psi^h_{\ ,j})=\varepsilon_{ij}+\xi_{ij}.
\end{equation}
For convenience, we have introduced the notation $\varepsilon_{ij}=\frac{1}{2}(\Psi_{i,j}+\Psi_{j,i})$ for the linear stretching strains and $\xi_{ij}=\frac{1}{2}(\Psi_{h,i}\Psi^h_{\ ,j})$ for the nonlinear stretching strains. Next we expand the potential of the external loads as 
\begin{equation}
P[\vett{\Psi(\vett{x})}]=P_1[\vett{\Psi(\vett{x})}]+P_2[\vett{\Psi(\vett{x})}]+\ldots,
\end{equation}
where $P_1[\vett{\Psi(\vett{x})}]$ is linear in $\vett{\Psi}$, and $P_2[\vett{\Psi(\vett{x})}]$ is quadratic in $\vett{\Psi}$, and so on. We now write the potential energy as
\begin{equation}
\label{eqVfull}
\mathcal{V}[\vett{\Psi(\vett{x})}]=\int_V\left[S^{ij}(\varepsilon_{ij}+\xi_{ij})+\frac{1}{2}\left(\frac{\partial^2 W}{\partial \gamma_{ij} \partial \gamma_{kl}}\right)_I\gamma_{ij}\gamma_{kl}+\ldots\right] \dd V +P_1[\vett{\Psi(\vett{x})}]+P_2[\vett{\Psi(\vett{x})}]+\ldots.
\end{equation}
Since the fundamental state is an equilibrium state, the first variation of the potential energy must be stationary for all admissible displacement displacement fields. The first variation contains only the leading order, linear terms of equation~\ref{eqVfull}, therefore ~\cite{koiter, thompsonRed, lanczos}
\begin{equation}
\label{eqd1V}
\delta\mathcal{V}[\vett{\Psi(\vett{x})}]=\int_V S^{ij}\varepsilon_{ij} \ \dd V +P_1[\vett{\Psi(\vett{x})}]=0.
\end{equation}
Accounting for equation~\ref{eqd1V}, the total potential energy becomes
\begin{equation}
\label{eqd1Ve}
\mathcal{V}[\vett{\Psi(\vett{x})}]=\int_V \left[S^{ij}\xi_{ij}+\frac{1}{2}\left(\frac{\partial^2 W}{\partial \gamma_{ij} \partial \gamma_{kl}}\right)_I\gamma_{ij}\gamma_{kl}\right] \dd V + P_2[\vett{\Psi(\vett{x})}].
\end{equation}
Since the primary focus of this paper will be to address the buckling of a spherical shell under the combined loading of pressure and curvature, it will be adequate to follow the approach of Koiter and limit our focus to {\em dead} loading. Therefore, we will only retain the potential of external loads that are linear in the displacement field, {\em i.e.} we will retain $P_1[\vett{\Psi(\vett{x})}]$ which is necessary for establishing equilibrium through equation~\ref{eqd1V}, and will neglect $P_2[\vett{\Psi(\vett{x})}]$, allowing us to write the total potential energy as:
\begin{equation}
\label{eqd1Vf}
\mathcal{V}[\vett{\Psi(\vett{x})}]=\int_V \left[S^{ij}\xi_{ij}+\frac{1}{2}\left(\frac{\partial^2 W}{\partial \gamma_{ij} \partial \gamma_{kl}}\right)_I\gamma_{ij}\gamma_{kl}\right] \dd V.
\end{equation}
As is typical in the theory of elasticity, the second derivative of the elastic potential per unit volume is defined as the tensor of elastic moduli
\begin{equation}
\label{eqAfund}
\mathcal{A}^{ijkl}|_I\equiv\left(\frac{\partial^2 W}{\partial \gamma_{ij} \partial \gamma_{kl}}\right)_I \\ 
\end{equation}
Here we have written the elastic moduli tensor evaluated in the fundamental state as opposed to the undeformed state, which is typically used in the theory of elasticity. The tensor of elastic moduli for a homogenous, isotropic material is given as:
\begin{equation}
\label{eqA3D}
\mathcal{A}^{ijkl}= G\left(\accentset{\circ}{\cg}^{ik}\accentset{\circ}{\cg}^{jl}+\accentset{\circ}{\cg}^{il}\accentset{\circ}{\cg}^{jk}+\frac{2\nu}{1-\nu}\accentset{\circ}{\cg}^{ij}\accentset{\circ}{\cg}^{kl}\right),
\end{equation}
where $G$ is the shear modulus, $\nu$ is Poisson's ratio, and $\accentset{\circ}{\cg}^{ij}$ denotes the contravariant components of the three dimensional metric tensor, with the overcircle decoration on a variable denoting a parameter in the undeformed, reference configuration. This approximation introduces an error of $\mathcal{O}(\epsilon)$, which corresponds to the difference in the metric tensors in the fundamental state $\cg^{ij}|_I$ and the undeformed state described by Kronecker's delta $\delta^{ij}$, {\em i.e.} $\cg^{ij}|_I-\delta^{ij}=\mathcal{O}(\epsilon)$, where $\epsilon$ is the largest principal extension in the fundamental state~\cite{koiter}. In what follows, we will use equation~\ref{eqA3D}, as the relative error $\mathcal{O}(\epsilon)$ introduced will not affect the positive--definite character of the potential energy~\cite{koiter}. Following equations~\ref{eqd1Vf} and ~\ref{eqA3D}, let us define the strain energy per unit volume of the undeformed body $\Phi$ and the pre--stress $\Upsilon$ as
\begin{subequations}
\begin{align}
\label{Phi}
\Phi &= \frac{1}{2}A^{ijkl}\gamma_{ij}\gamma_{kl}, \\
\label{Up}
\Upsilon &= S^{ij}\xi_{ij},
\end{align}
\end{subequations}
such that we can write the total potential energy as
\begin{equation}
\label{V3D}
\mathcal{V}=\int_V \left[\Upsilon+\Phi\right] \dd V. 
\end{equation}
 
According to the general theory of elastic stability, this equilibrium state is unstable if the potential energy functional is indefinite. A positive--definite energy functional requires that $\delta^2\mathcal{V}+\delta^3\mathcal{V}+\delta^4\mathcal{V} \geq 0$. The third and fourth variations only contain higher--order terms, and so it is usually sufficient to restrict our attention to the character of the second variation, {\em i.e.} $\delta^2\mathcal{V}\geq0$. The stability limit for a material that follows the generalized Hooke's law which is subjected to conservative, dead--weight loading is determined by the character of the second variation of its potential energy functional, $\delta^2\mathcal{V}$. By virtue of the calculus of variations, the second variation will only contain terms of second order in the displacements~\cite{lanczos, thompsonRed}, which corresponds to terms containing $\xi_{ij}$ and the linear terms in the product $\gamma_{ij}\gamma_{kl}$, {\em i.e.} $\varepsilon_{ij}\varepsilon_{kl}$. These terms can be readily identified from equation~\ref{eqd1Ve}, leading to the stability criteria
\begin{equation}
\label{eqd2V}
\delta^2\mathcal{V}[\vett{\Psi(\vett{x})}]=\int_V \left[S^{ij}\xi_{ij}+\frac{1}{2}\mathcal{A}^{ijkl}\varepsilon_{ij}\varepsilon_{kl}\right] \dd V > 0.
\end{equation}
The character of equation~\ref{eqd2V} will determine the stability of the shells under investigation. Equivalently, we may check to see if the third variation is stationary, which is done in \S\ref{stabilityAnalysis}.

\section{Shell Theory -- Strain Energy of an Deformed State}
\label{shelltheory}
The stability of thin elastic shells subjected to a combination of pressure and curvature loading is an inherently three--dimensional problem, however great strides can be made by reducing the dimensionality to two--dimensional problem that is posed on the middle--surface of the shell.  This will require us to approximate the total potential energy of the three--dimensional material given by equation~\ref{eqd2V} with a two--dimensional energy. As noted above, we will assume the material is elastic, homogenous, and isotropic. Additionally, we will assume that the states of stress is approximately plane and parallel to the middle surface, and that the strains are small everywhere. These assumptions are effectively equivalent to the mutually contradictory Kirchhoff--Love assumptions, and they enable us to write the elastic energy of the shell as the sum of stretching and bending energies, as was shown qualitatively by Koiter~\cite{Koiter1960}, later rigorously proven by John~\cite{John1965}, and discussed again recently by Efrati~\cite{Efrati2009}. With the approximation of plane stress~\footnote{The assumption that the shell is in a state of approximate plane stress may in fact be omitted, as it is a consequence of the small strain assumption~\cite{John1965}.}, the transverse shear strains are zero, $\gamma_{\alpha 3}=0$, and the transverse normal strain is given by
\begin{equation}
\label{gamma33}
\gamma_{33}=-\frac{\nu}{1-\nu}\accentset{\circ}{\cg}^{\alpha\beta}\gamma_{\alpha\beta}.
\end{equation}
To make additional progress we need to evaluate $\accentset{\circ}{\cg}^{\alpha\beta}$, and we will accomplish this by investigating the geometry of the middle surface of the shell. 

\begin{figure}[t]
\begin{center}
\includegraphics[width=0.5\columnwidth]{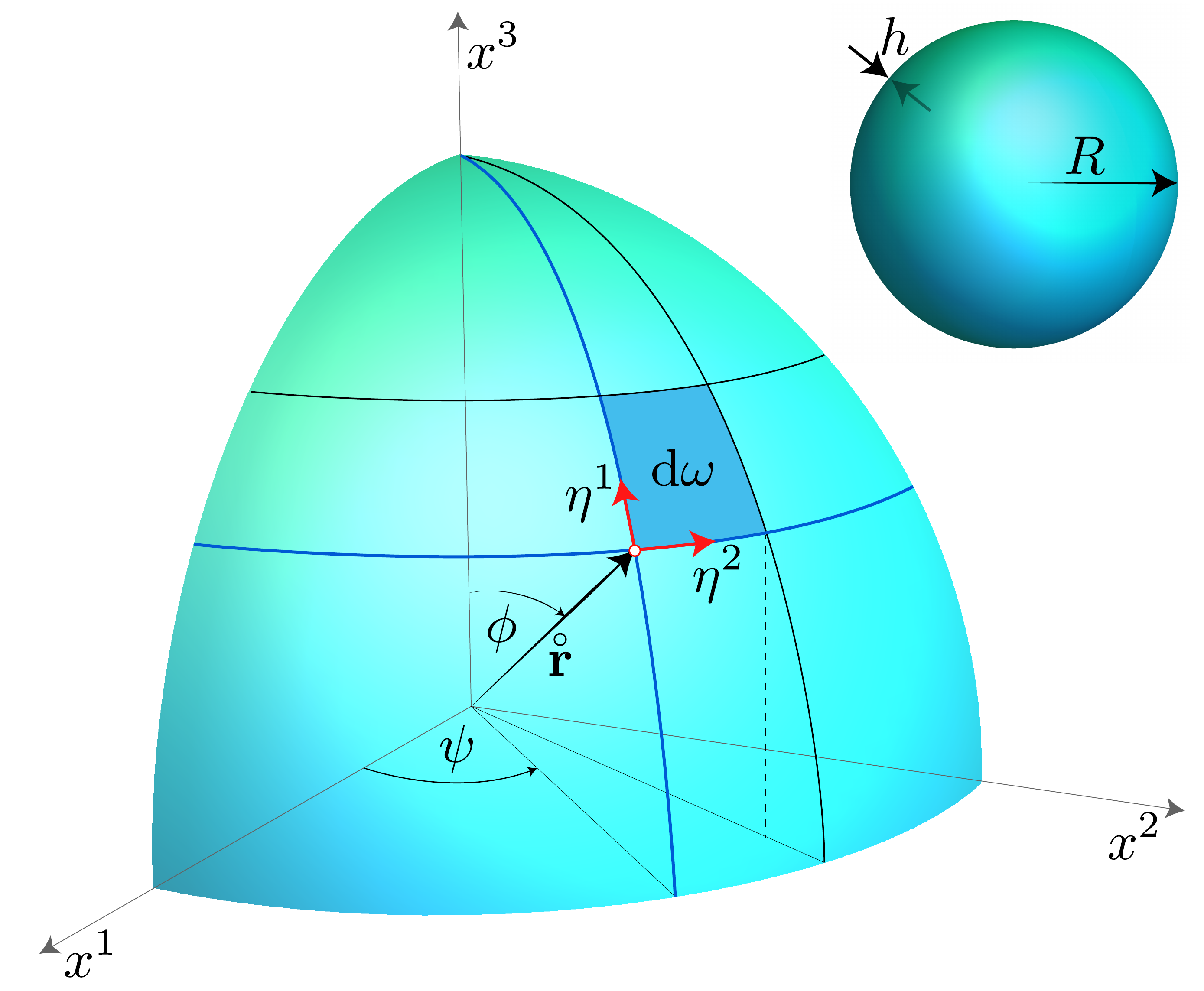}
\caption{\label{scheme} A schematic of a spherical shell (inset) along with a shell segment denoting the curvilinear coordinates $\eta^\alpha$ at a point a distance $\protect\accentset{\circ}{\vett{r}}$ from some origin in space.}
\end{center}
\end{figure}

A point in the three--dimensional space of shell will be identified from its distance $\eta^3$ to the middle surface, and by the surface coordinates of its projection onto the middle surface. If the shell has a constant thickness $h$, the inner and outer faces of the shell, {\em i.e.} $\eta^3=z=\pm \frac{1}{2}h$, will be parallel to the mid--surface. Formally, we can say that our two--dimensional surface $\mathcal{S}=\accentset{\circ}{\bm{r}}(x^1, x^2)$ is embedded in $\mathds{R}^3$, and is parameterized by $y=(\eta^1,\eta^2)$. With this parameterization we can define the covariant tangent vectors to the surface as 
\begin{equation}
\label{eq-aVec}
\accentset{\circ}{\bm{a}}_\alpha=\accentset{\circ}{\bm{r}}_{,\alpha}\equiv \frac{\partial \accentset{\circ}{\bm{r}}}{\partial \eta^\alpha}.
\end{equation}
The {\em first fundamental form} of the surface is determined by by the distance $\dd s$ between two neighboring points on the surface
\begin{equation}
\dd \accentset{\circ}{s}^2=\accentset{\circ}{\bm{a}}_\alpha\cdot\accentset{\circ}{\bm{a}}_\beta \ \dd\eta^\alpha\eta^\beta=\accentset{\circ}{a}_{\alpha\beta}\dd\eta^\alpha\eta^\beta,
\end{equation}
where $\accentset{\circ}{a}_{\alpha\beta}$ is the metric tensor of the surface. The metric tensor is symmetric such that $\accentset{\circ}{a}_{\alpha\beta}=\accentset{\circ}{a}_{\beta\alpha}$, and the inverse metric is defined by $\accentset{\circ}{a}^{\alpha\gamma}\accentset{\circ}{a}_{\gamma\beta}=\delta_\beta^\alpha$. The metric tensor of the middle surface contains all information about lateral distance between points on the mid--surface of the shell. To quantify how the shell curves as you move between points on the mid--surface, we can project a vector normal $\accentset{\circ}{\bm{n}}$ to the surface at a given point onto the metric tensor. The unit--length normal vector is defined by 
\begin{equation}
\label{eq-n}
\accentset{\circ}{\bm{n}}\equiv\accentset{\circ}{\bm{a}}^3=\frac{\accentset{\circ}{\bm{a}}_1 \times \accentset{\circ}{\bm{a}}_2}{|\accentset{\circ}{\bm{a}}_1 \times \accentset{\circ}{\bm{a}}_2|}.
\end{equation}
Projecting this vector onto the surface enables us to construct a covariant tensor of second order, $\accentset{\circ}{b}_{\alpha\beta}=\accentset{\circ}{\bm{n}}\cdot\accentset{\circ}{\bm{a}}_{\alpha,\beta}$, which measures local curvature on the middle surface of the shell. More formally, the distance of a point on the surface of the shell to a plane tangent to a nearby origin is given by the {\em second fundamental form}
\begin{equation}
\accentset{\circ}{\mathrm{I\!I}}=\frac{1}{2}\accentset{\circ}{b}_{\alpha\beta}\dd\eta^\alpha\dd\eta^\beta,
\end{equation}
where $\accentset{\circ}{b}_{\alpha\beta}$ are the covariant coefficients of the second fundamental form; with a mild abuse of terminology, we will refer to $\accentset{\circ}{b}_{\alpha\beta}$ as the curvature tensor of the surface. With the definitions of the first and second fundamental forms in hand, we can define the two surface invariants of the shell -- the mean curvature $\accentset{\circ}{\mathcal{H}}$ and the Gaussian curvature $\accentset{\circ}{\mathcal{K}}$ as
\begin{subequations}
\begin{align}
\label{H}
\accentset{\circ}{\mathcal{H}}&=\frac{1}{2}\accentset{\circ}{a}^{\alpha\beta}\accentset{\circ}{b}_{\alpha\beta}=\frac{1}{2}\accentset{\circ}{b}_\alpha^\alpha, \\
\label{K}
\accentset{\circ}{\mathcal{K}}&=\frac{\accentset{\circ}{b}}{\accentset{\circ}{a}}=\accentset{\circ}{b}_1^1\accentset{\circ}{b}_2^2-\accentset{\circ}{b}_2^1\accentset{\circ}{b}_1^2,
\end{align}
\end{subequations}
where the determinant is introduced by $\accentset{\circ}{a}=|\accentset{\circ}{a}_{\alpha\beta}|=\accentset{\circ}{a}_{11}\accentset{\circ}{a}_{22}-(\accentset{\circ}{a}_{12})^2$ and $\accentset{\circ}{b}=|\accentset{\circ}{b}_{\alpha\beta}|=\accentset{\circ}{b}_{11}\accentset{\circ}{b}_{22}-(\accentset{\circ}{b}_{12})^2$~\cite{Oneill1997}. Returning to equation~\ref{gamma33}, we can now specify the covariant components of the spatial metric tensor $\accentset{\circ}{\cg}_{ij}$ in terms of the metric tensor and curvature tensor of the middle surface by expanding through the thickness with a Taylor series, such that
\begin{subequations}
\begin{align}
\label{gab}
\accentset{\circ}{\cg}_{\alpha\beta}&=\accentset{\circ}{a}_{\alpha\beta}-2z\accentset{\circ}{b}_{\alpha\beta}+z^2\accentset{\circ}{b}_\alpha^\kappa \accentset{\circ}{b}_{\kappa\beta}, \\ 
\label{g1312}
\accentset{\circ}{\cg}_{13}&=\accentset{\circ}{\cg}_{12}=0, \\
\label{g33}
\accentset{\circ}{\cg}_{33}&=1.
\end{align}
\end{subequations}
We are now in position to reduce the integral over the volume of the shell given by equation~\ref{V3D} into an integral over the shell's surface area. We note that the volume element of the shell is given by
\begin{equation}
\label{dV}
\dd \accentset{\circ}{V}=\sqrt{\accentset{\circ}{\cg}} \ \dd \eta^1 \dd \eta^2 \dd z =\sqrt{\accentset{\circ}{a}} \ \dd \eta^1 \dd \eta^2 \sqrt{\frac{\accentset{\circ}{\cg}}{\accentset{\circ}{a}}} \ \dd z,
\end{equation} 
where, using equations~\ref{H},~\ref{K}, and~\ref{gab} we can write
\begin{equation}
\label{detgdeta}
\sqrt{\frac{\accentset{\circ}{\cg}}{\accentset{\circ}{a}}}=1-2z\accentset{\circ}{\mathcal{H}}+z^2\accentset{\circ}{\mathcal{K}}.
\end{equation}
Inserting equation~\ref{detgdeta} into equation~\ref{Phi} allows us to write the strain energy per unit volume of the undeformed shell as
\begin{equation}
\label{VPhi}
\mathcal{U}[\vett{\Psi(\vett{\eta})}]= \int_\mathcal{S} \sqrt{\accentset{\circ}{a}} \ \dd \eta^1 \dd \eta^2\int_{-h/2}^{h/2}\left(1-2z\accentset{\circ}{\mathcal{H}}+z^2\accentset{\circ}{\mathcal{K}}\right)\Phi[\eta^\alpha,z] \ \dd z
\end{equation}
The strain energy density $\Phi[\eta^\alpha, z]$ can be expanded in a Taylor series with respect to the coordinate normal to the middle surface
\begin{equation}
\label{PhiTaylor}
\Phi[\eta^\alpha, z] = \Phi[\eta^\alpha, 0]+z \nabla_3 \Phi[\eta^\alpha, 0] +\frac{z^2}{2} \nabla_{33} \Phi[\eta^\alpha, 0]+\ldots, 
\end{equation}
where $\nabla_i$ represents covariant differentiation with respect to $\eta^i$, and $\nabla_{\alpha\beta}$ is the second order covariant differential operator. The covariant derivatives of the elastic moduli tensor vanish, leading us to write
\begin{equation}
\label{PhiA}
\Phi[\eta^\alpha, z] = \frac{1}{2}\mathcal{A}^{ijkl}\left(\gamma_{ij}+z\nabla_3\gamma_{ij}+\frac{z^2}{2}\nabla_{33}\gamma_{ij}+\ldots\right)\left(\gamma_{kl}+z\nabla_3\gamma_{kl}+\frac{z^2}{2}\nabla_{33}\gamma_{kl}+\ldots\right)
\end{equation}
With equation~\ref{PhiA} we can now integrate the right hand side of equation~\ref{VPhi} over the thickness. Consistent with our requirement of small strains, we can retain only the first two terms following integration. Justification for the neglecting these higher order terms was given by Koiter~\cite{Koiter1960}, and later quantified by John~\cite{John1965}. Here, we apply the contradictory assumption of plane strain, thus setting
\begin{equation}
\gamma_{33} =0,
\end{equation}
which reduces the approximate strain energy per unit area to 
\begin{equation}
\mathcal{U}=\frac{h}{2}\mathcal{A}^{\alpha\beta\lambda\mu}\gamma_{\alpha\beta}\gamma_{\lambda\mu} +\frac{h^3}{24}\mathcal{A}^{\alpha\beta\lambda\mu}\nabla_3\gamma_{\alpha\beta}\nabla_3\gamma_{\lambda\mu},
\end{equation}
where the assumption of a plane state of stress resulted in a decoupling of the membrane or stretching energy density which is linear in $h$ and the bending energy density which is preceded by $h^3$. In this approximate strain energy per unit area, the tensor of elastic moduli on the middle surface is given by
\begin{equation}
\label{Ashell}
\mathcal{A}^{\alpha\beta\lambda\mu}=\frac{E}{2(1+\nu)}\left[\accentset{\circ}{a}^{\alpha\lambda}\accentset{\circ}{a}^{\beta\mu}+\accentset{\circ}{a}^{\alpha\mu}\accentset{\circ}{a}^{\beta\lambda}+\frac{2\nu}{1-\nu}\accentset{\circ}{a}^{\alpha\beta}\accentset{\circ}{a}^{\lambda\mu}\right].
\end{equation}
Natural choices for the middle surface strain tensor and the tensor of changes of curvature are
\begin{subequations}
\begin{align}
\label{gammaForm}
\gamma_{\alpha\beta}&=\frac{1}{2}(a_{\alpha\beta}-\accentset{\circ}{a}_{\alpha\beta}), \\
\label{rhoForm}
\varrho_{\alpha\beta}&=b_{\alpha\beta}-\accentset{\circ}{b}_{\alpha\beta},
\end{align}
\end{subequations}
where $a_{\alpha\beta}$ and $b_{\alpha\beta}$ represent the metric tensor and curvature tensor of the deformed shell, respectively. Finally, we can write the strain energy per unit area as
\begin{equation}
\mathcal{U}=\frac{h}{2}\mathcal{A}^{\alpha\beta\lambda\mu}\gamma_{\alpha\beta}\gamma_{\lambda\mu} +\frac{h^3}{24}\mathcal{A}^{\alpha\beta\lambda\mu}\varrho_{\alpha\beta}\varrho_{\lambda\mu}.
\end{equation}
We note that if $E$ and $h$ are homogenous, the tensor of elastic moduli can greatly simplify the strain energy of the shell into the familiar form
\begin{equation}
\label{UpostA}
\mathcal{U}=  \frac{Y}{2} \int \Bigl[(1-\nu)\gamma^{\alpha\beta}\gamma_{\alpha\beta}+\nu(\gamma_\alpha^\alpha)^2\Bigr] \dd\omega+  \frac{B}{2}  \int \Bigl[(1-\nu)\varrho^{\alpha\beta}\varrho_{\alpha\beta}+\nu(\varrho_\alpha^\alpha)^2\Bigr] \dd\omega,
\end{equation}
with stretching and bending rigidities given by $Y=\frac{Eh}{(1-\nu^2)} $ and $B=\frac{Eh^3}{12(1-\nu^2)}$, respectively.

\section{Shell Theory -- Strain Energy in the Fundamental State}
\label{fundamental}

Stability of the shell must be evaluated in the {\em fundamental state} -- a deformed configuration that retains the shell's initial symmetries.  In the fundamental state, the shell has deformed in response to a combination of internal and externally applied forces and moments, represented by the pre--stress $\Upsilon$ in equation~\ref{Up}. We will next reduce the contribution of $\Upsilon$ to the membrane and curvature stresses on the middle surface of the shell. From the strain energy we can define the stress resultants and stress couples as partial derivatives of the strain energy per unit area of the middle surface with respect to the middle surface strains and the changes of curvature~\cite{Koiter1960}. Therefore, the corresponding symmetric contravariant tensors are given by
\begin{subequations}
\begin{align}
\label{NabO}
N^{\alpha\beta} &= \frac{\partial \mathcal{U}}{\partial \gamma_{\alpha\beta}}=h\mathcal{A}^{\alpha\beta\lambda\mu}\gamma_{\lambda\mu}=\frac{Eh}{2(1+\nu)}\left(\accentset{\circ}{a}^{\alpha\lambda}\accentset{\circ}{a}^{\beta\mu}+\frac{\nu}{1-\nu}\accentset{\circ}{a}^{\alpha\beta}\accentset{\circ}{a}^{\lambda\mu}\right)(a_{\lambda\mu}-\accentset{\circ}{a}_{\lambda\mu}), \\
\label{MabO}
M^{\alpha\beta} &= \frac{\partial \mathcal{U}}{\partial \varrho_{\alpha\beta}}=\frac{h^3}{12}\mathcal{A}^{\alpha\beta\lambda\mu}\varrho_{\lambda\mu}=\frac{Eh^3}{12(1+\nu)}\left(\accentset{\circ}{a}^{\alpha\lambda}\accentset{\circ}{a}^{\beta\mu}+\frac{\nu}{1-\nu}\accentset{\circ}{a}^{\alpha\beta}\accentset{\circ}{a}^{\lambda\mu}\right)(b_{\lambda\mu}-\accentset{\circ}{b}_{\lambda\mu}).
\end{align}
\end{subequations}
From Hooke's law, we can write $S^{\alpha\beta}=\mathcal{A}^{\alpha\beta\lambda\mu}\gamma_{\lambda\mu}$. Using the definition of $\accentset{\circ}{\cg}_{\alpha\beta}$ from equation~\ref{gab} and the membrane and curvature stresses given by equations~\ref{NabO} and~\ref{MabO}, respectively we can rewrite the second Piola--Kirchhoff stress tensor on the middle surface as~\cite{koiter}
\begin{equation}
\label{Sab}
S^{\alpha\beta}=\frac{1}{h}N^{\alpha\beta}-12\frac{z}{h^3}M^{\alpha\beta}.
\end{equation}
From equation~\ref{Up}, we note that $S^{\alpha\beta}$ multiplies only the nonlinear terms in the Green--Lagrange strain tensor, {\em i.e.} $\xi_{ij}=\frac{1}{2}(\Psi_{h,i}\Psi^h_{\ ,j})$. On the middle surface of the shell the partial derivatives in $\xi$ are replace with covariant derivatives, and we can write
\begin{equation}
\label{xiab}
\xi_{\alpha\beta}\equiv\frac{1}{2}\nabla_\alpha \Psi^\kappa\nabla_\beta \Psi_\kappa = \gamma_{\alpha\beta}-\frac{1}{2}\left(\accentset{\circ}r^\kappa_{,\alpha}\nabla_\beta \Psi_\kappa+\accentset{\circ}r_{\kappa, \beta}\nabla_\alpha\Psi^\kappa\right).
\end{equation}
Now we can evaluate the stress in the fundamental state as
\begin{equation}
\label{Upab}
\Upsilon=\frac{1}{2}\int_V S^{\alpha\beta}\nabla_\alpha \Psi^\kappa\nabla_\beta \Psi_\kappa \dd V = \int_\mathcal{S} \sqrt{a} \ \dd x^1 \dd x^2 \int_{-h/2}^{h/2}\left[\frac{1}{h}N^{\alpha\beta}-12\frac{z}{h^3}M^{\alpha\beta}\right]\left(\xi_{\alpha\beta}-z\zeta_{\alpha\beta}\right) \dd z, 
\end{equation}
where $\zeta_{\alpha\beta}$ are the nonlinear bending strains. Finally, recalling equation~\ref{eqd1Vf}, we can write the total potential energy in the fundamental state as
\begin{equation}
\label{V2D}
\mathcal{V}[\vett{\Psi}]=\int\left[N^{\alpha\beta}\xi_{\alpha\beta}+M^{\alpha\beta}\zeta_{\alpha\beta}+\frac{h}{2}\mathcal{A}^{\alpha\beta\lambda\mu}\gamma_{\alpha\beta}\gamma_{\lambda\mu} +\frac{h^3}{24}\mathcal{A}^{\alpha\beta\lambda\mu}\varrho_{\alpha\beta}\varrho_{\lambda\mu}\right]\dd\omega,
\end{equation}
where we have introduced $\dd \omega = \sqrt{a} \ \dd \eta^1\eta^2$. Utilizing equation~\ref{Ashell}, we may rewrite equation~\ref{V2D} as
\begin{equation}
\label{VpostA}
\mathcal{V}[\vett{\Psi}]= \int \frac{Eh}{2(1-\nu^2)}\Bigl[(1-\nu)\gamma^{\alpha\beta}\gamma_{\alpha\beta}+\nu(\gamma_\alpha^\alpha)^2+\frac{h^2}{12}[(1-\nu)\varrho^{\alpha\beta}\varrho_{\alpha\beta}+\nu(\varrho_\alpha^\alpha)^2]\Bigr]+N^{\alpha\beta}\xi_{\alpha\beta}+M^{\alpha\beta}\zeta_{\alpha\beta} \ \dd\omega.
\end{equation}

It is at this point that we will deviate from the classical treatment of the stability of thin, elastic shells. It is common to simplify equation~\ref{V2D} further by making two assumptions (1.) linear curvature strains are sufficient to characterize the strain energy of the shell in the fundamental state, and (2.) the curvature stress $M^{\alpha\beta}$ can be neglected. Indeed, we find the validation for these approximations in reference~\cite{Koiter1967}, where Koiter writes:
\begin{myindentpar}{18mm}{18mm}
\textit{``If we restrict our attention to fundamental states I in which bending stresses do not exceed the membrane stresses in order of magnitude, we may therefore presumably neglect the nonlinear terms in the changes in curvature. The restriction implied by this simplification is not at all serious. We are not aware of any significant shell buckling problem in which the fundamental state involves membrane stresses which are small in comparison with the bending stresses. Moreover, if we restrict out attention to ``small finite deflections'' in the sense of~\cite{Koiter1966}, the changes of curvature may always be represented by their linear approximation without any loss in accuracy within the framework of shell theory. Finally, the most important shell buckling problems are those in which the fundamental state I is (approximately) a membrane state of stress, and the tensor resultants $M^{\alpha\beta}$ may then be neglected everywhere.''}~\cite{Koiter1967}
\end{myindentpar}
Recent work by the authors have demonstrated that retaining the contribution from the curvature stress is essential when analyzing the buckling of thin elastic shells subjected to an evolving natural curvature~\cite{Pezzulla2018}.  Neglecting $M^{\alpha\beta}\zeta_{\alpha\beta}$ when calculating the critical buckling curvature leads to a result that is erroneous in magnitude. Since this result confirms that bending stresses are at least of the same order of magnitude of the membrane stresses, we will also retain the nonlinear curvature strains in equation~\ref{V2D}. The presence of these higher order terms will facilitate the investigation of the stability of the critical points, through the character of the third variation $\delta^3\mathcal{V}$ which the authors will leave to future work. Retaining all the terms in equation~\ref{V2D}, we can now turn to the question of stability in the presence of both an externally applied pressure and an internally varying natural curvature. 

From the theory of elastic stability, we know that the character of the second variation of the total potential energy will characterize the stability of the system. Recalling that we write the stretching strain $\gamma_{\alpha\beta}$ as the superposition of the linear strains $\varepsilon_{\alpha\beta}$ and the nonlinear stretching strains $\xi_{\alpha\beta}$, we will similarly write the curvature strains $\varrho_{\alpha\beta}$ as the superposition of the linear curvature strains $\rho_{\alpha\beta}$ and the nonlinear curvature strains $\zeta_{\alpha\beta}$. Therefore, utilizing equation~\ref{eqd2V}, the second variation of $\mathcal{V}$ is given by
\begin{equation}
\label{d2V}
\delta^2\mathcal{V}=\int \frac{Eh}{2(1-\nu^2)}\Bigl[(1-\nu)\varepsilon^{\alpha\beta}\varepsilon_{\alpha\beta}+\nu(\varepsilon_\alpha^\alpha)^2+\frac{h^2}{12}[(1-\nu)\rho^{\alpha\beta}\rho_{\alpha\beta}+\nu(\rho_\alpha^\alpha)^2]\Bigr]+N^{\alpha\beta}\xi_{\alpha\beta}+M^{\alpha\beta}\zeta_{\alpha\beta}\ \dd\omega,
\end{equation}
where we split the second variation into contributions from the strain energy in the shell $\delta^2\mathcal{V}_0$, the energetic contribution from the membrane prestress $\delta^2\mathcal{V}_N$, and the bending prestress $\delta^2\mathcal{V}_M$. In the subsequent analysis, we will closely follow Koiter's treatment of the buckling behavior of a spherical shell under uniform external pressure, which was recently collected as part of a set of lecture notes on elastic stability~\cite{koiter}. While we are indebted to these notes, we caution the reader that there are numerous typographical errors in the analysis, which we have done our best to remedy here. Our addition to Koiter's classical treatment is the incorporation of a stimulus that induces a change in the shell's natural curvature. As we will see, the natural curvature contributes to both the membrane and bending stress in the fundamental state, and has a significant affect on the classical result for the critical buckling pressure. The result of the forthcoming analysis will be the critical buckling pressure of a shell exposed to both uniform external pressure $p$ and an evolving natural curvature $\kappa$. 

We outline our approach to identifying the critical buckling pressure as: (1.) Write the stretching and bending strains in equation~\ref{d2V} in terms of the displacement vector $\vett{\Psi}$, and apply these general equations to a complete spherical shell, (2.) we will split the tangential displacement field into two invariants that will aid in simplifying the functional given by equation~\ref{d2V}, (3.) we will then expand the two remaining displacement vectors in a series of spherical surface harmonics, and finally (4.) perform linear stability analysis and determine the critical buckling pressure by noting that the wavenumber of the buckling pattern will be large.

We begin with the strain fields $\gamma_{\alpha\beta}$ and $\varrho_{\alpha\beta}$ defined by equations~\ref{gammaForm} and~\ref{rhoForm}, respectively. It is helpful to define the two--dimensional deformation gradient $\chi_{\alpha\beta}$, the rotation in the tangent plane to the shell $\omega_{\alpha\beta}$, and the rotation of the normal vector $\accentset{\circ}{\vett{n}}$ as $\varphi_{\alpha}$ in terms of the displacement vector $\vett{\Psi}$ as~\cite{Niordson1985, koiter}
\begin{subequations}
\begin{align}
\label{chi}
\chi_{\alpha\beta} &= \nabla_\alpha \Psi_\beta-\accentset{\circ}{b}_{\alpha\beta}\Psi_3,\\
\label{omega}
\omega_{\alpha\beta} &= \frac{1}{2}\left(\nabla_\beta \Psi_\alpha- \nabla_\alpha \Psi_\beta \right),\\
\label{varphi}
\varphi_{\alpha} &= \Psi_{3,\alpha}+\accentset{\circ}{b}_\alpha^\kappa\Psi_\kappa.
\end{align}
\end{subequations}
We can now write the strain tensor of the middle surface $\gamma_{\alpha\beta}$ and the curvature strain tensor $\varrho_{\alpha\beta}$ as~\cite{Niordson1985} as~\cite{Niordson1985, koiter}
\begin{subequations}
\begin{align}
\label{gammaab}
\gamma_{\alpha\beta}&=\frac{1}{2}\left(\nabla_\beta \Psi_\alpha+ \nabla_\alpha \Psi_\beta \right) -\accentset{\circ}{b}_{\alpha\beta}\Psi_3+\frac{1}{2}\left(\varepsilon_{\alpha}^\lambda-\omega_{\alpha}^\lambda\right)\left(\varepsilon_{\lambda\beta}-\omega_{\lambda\beta}\right) +\frac{1}{2}\varphi_\alpha\varphi_\beta, \\
\label{varrhoab}
\varrho_{\alpha\beta}&=\sqrt{\frac{\accentset{\circ}{a}}{a}}\left[\left(1+\chi_\lambda^\lambda+\frac{\chi}{\accentset{\circ}{a}}\right)\left(\accentset{\circ}{b}_{\alpha\beta}+\nabla_\beta\varphi_\alpha+\accentset{\circ}{b}_\beta^\gamma\chi_{\alpha\gamma}\right)-\left(\varphi^\mu+\epsilon^{\mu\eta}\epsilon^{\gamma\delta}\varphi_\gamma\chi_{\delta\eta}\right)\left(\nabla_\beta\chi_{\alpha\beta}-\accentset{\circ}{b}_{\beta\mu}\varphi_\alpha \right)\right]-\accentset{\circ}{b}_{\alpha\beta},
\end{align}
\end{subequations}
where $\epsilon^{\mu\eta}$ is introduced as the two--dimensional Levi-Civita permutation tensor, and the determinant of $\chi_{\alpha\beta}$ is again written as $\chi=|\chi|$. The linear parts of the stretching and curvature strains are then identified as
\begin{subequations}
\begin{align}
\label{ep}
\varepsilon_{\alpha\beta} &= \frac{1}{2}\left(\nabla_\beta \Psi_\alpha+ \nabla_\alpha \Psi_\beta \right) -\accentset{\circ}{b}_{\alpha\beta}\Psi_3,\\
\label{rho}
\rho_{\alpha\beta} &= \nabla_\beta\varphi_\alpha+\accentset{\circ}{b}_\beta^\gamma\chi_{\alpha\gamma}.
\end{align}
\end{subequations}

It is at this point that we apply these results to the specific case of a spherical shell under uniform pressure and a homogenous nature curvature. The basic property of a spherical shell is that the first and second fundamental tensors are proportional to each other through the radius of curvature of the shell's middle surface, $R$, {\em i.e.}
\begin{equation}
\label{bS}
\accentset{\circ}{b}_{\alpha\beta}=-\frac{1}{R}\accentset{\circ}{a}_{\alpha\beta}
\end{equation}
The negative sign used in equation~\ref{bS} is chosen such that the positive direction of the normal vector $\accentset{\circ}{\vett{n}}$ points outward. This sign convention is consistent with the work of Niordson~\cite{Niordson1985}, but opposite of what was used by Koiter~\cite{koiter1969}. With this definition in hand, we can rewrite equations~\ref{ep}, and~\ref{rho} as
\begin{subequations}
\begin{align}
\label{varphiS}
\varphi_{\alpha} &= \Psi_{3,\alpha}-\frac{1}{R}\Psi_\alpha. \\
\label{epS}
\varepsilon_{\alpha\beta} &= \frac{1}{2}\left(\nabla_\beta \Psi_\alpha+ \nabla_\alpha \Psi_\beta \right)+\frac{1}{R}\accentset{\circ}{a}_{\alpha\beta}\Psi_3,\\
\label{rhoS}
\rho_{\alpha\beta} &= \nabla_{\alpha\beta}\Psi_{3}-\frac{2}{R}\varepsilon_{\alpha\beta}+\frac{1}{R^2}\accentset{\circ}{a}_{\alpha\beta}\Psi_3.
\end{align}
\end{subequations}

The potential of the uniform pressure is nonlinear because pressure acts on the change of volume of the shell. At leading order, the potential is linear in $\vett{\Psi}$, and this contribution is referred to as the {\em dead pressure}, which can be expressed as~\cite{koiter}
\begin{equation}
\label{Pp}
\mathcal{P}_p=p\int_\omega \Psi_3 \ \dd \omega.
\end{equation}
As discussed in section~\ref{stability}, considering only the dead--weight loading terms will be sufficient to establish the stability limit.  

The potential of the natural curvature was recently introduced by Pezzulla {\em et al.}~\cite{Pezzulla2018}, and is a stimulus that is conjugate to $\tr(b-\accentset{\circ}{b})$. In general, changes in the natural or spontaneous curvature of a thin shell may arise through differential swelling, heating, or growth, or may be induced, for example, by proteins within the cell membrane. As is common to problems in thermoelasticity, natural curvature will often leave the shell in a state of residual stress due to an incompatibility between the geometric changes prescribed by the stimulus and the ability for these geometries to be embedded in $\mathds{R}^3$. This discrepancy motivated theoretical advances aimed at modeling so--called {\em incompatible elasticity}, which introduced the multiplicative decomposition of the deformation gradient -- a concept stemming from models in plasticity theory -- to model the growth in three-dimensional elastic bodies~\cite{Kondaurov1987, Takamizawa1987}. These same ideas later motivated modeling incompatible plates and shells, through the so--called theory of {\em non--Euclidean plates and shells}~\cite{Efrati2009, Sadik2016, Pezzulla2017}. This reduced order model followed the same approximations Koiter used in developing the theory reproduced in section~\ref{shelltheory}, with the main distinction being the choice of stretching and bending strains of the middle surface. In these incompatible shell models, strains are measured with respect to the configuration that would make the shell stress--free as opposed to the initial, undeformed configuration of the shell. This introduces a new geometry of the middle surface, described by the first and second fundamental tensors $\overline{\vett{a}}$ and $\overline{\vett{b}}$. A key distinction between $\overline{\vett{a}}$ and $\overline{\vett{b}}$ and $\accentset{\circ}{\vett{a}}$ and $\accentset{\circ}{\vett{b}}$ is that $\overline{\vett{a}}$ and $\overline{\vett{b}}$ are generally not embeddable in Euclidean space. Therefore, new measures of stretching and bending strains are introduced as
\begin{subequations}
\begin{align}
\label{gammaNEP}
\overline{\gamma}_{\alpha\beta}&=\frac{1}{2}\left(a_{\alpha\beta}-\overline{a}_{\alpha\beta}\right), \\
\label{varrhoNEP}
\overline{\varrho}_{\alpha\beta}&=b_{\alpha\beta}-\overline{b}_{\alpha\beta}.
\end{align}
\end{subequations}
The choice of metric used to raise and lower indices for these incompatible shells is not necessarily obvious, as pointed out by Hanna~\cite{Hanna2017}, however the discrepancies that can emerge are important for determining the bending energy of a shell in the presence of middle surface stretching. For thin bodies undergoing differential swelling or heating through their thickness, the curvature inducing stimulus will not stretch the shell's middle surface~\cite{Pezzulla2018}, meaning $\overline{\vett{a}}=\accentset{\circ}{\vett{a}}$, and so we will continue to use the reference metric tensor to raise and lower indices. Since the stimulus does not stretch the middle surface, that enables an additive decomposition of the curvature tensor, in the form \cite{Pezzulla2017}
\begin{equation}
\label{bbar}
\overline{b}_{\alpha\beta}=\accentset{\circ}{b}_{\alpha\beta}+\kappa \accentset{\circ}{a}_{\alpha\beta}.
\end{equation}
As shown by Pezzulla {\em et al.}~\cite{Pezzulla2018}, the potential of the natural curvature can then be written as
\begin{equation}
\label{Pk}
\mathcal{P}_\kappa =-\frac{Eh}{2(1-\nu^2)}\int_\omega \frac{1+\nu}{6}h^2\kappa \varrho_\alpha^\alpha \ \dd \omega.
\end{equation}

Having identified the potentials of the uniform pressure and homogenous natural curvature, we can now derive the relationship between the displacement of the shell in response to curvature and pressure in the fundamental state. Deformations to the shell from uniform pressure or homogenous natural curvature that preserve the shell's spherical symmetry satisfy the conditions $\Psi_1=\Psi_2=0$, with $\Psi_3=f(p;\kappa)$. This makes it fairly straightforward to evaluate the relationship between normal displacement, pressure, and curvature. In the fundamental state, strains are small everywhere, and equation~\ref{epS} for $\varepsilon_{\alpha\beta}$ and~\ref{rhoS} for $\rho_{\alpha\beta}$ simplify significantly due to the constraints on $\Psi_\alpha$. As a result, and in combination with equations~\ref{Pp} and ~\ref{Pk}, we can write the total potential energy in the fundamental state as
\begin{equation}
\mathcal{V}[\Psi_3]\Big|_f=\frac{Eh}{2(1-\nu^2)}\int_\omega2(1+\nu)\left(\frac{\Psi_3}{R}\right)^2+\frac{1+\nu}{6}\frac{h^2\Psi_3^2}{R^4}+\frac{1+\nu}{3}\frac{h^2}{R^2}\kappa \Psi_3+\frac{2(1-\nu^2)}{Eh}p\Psi_3 \ \dd \omega
\end{equation}
If we denote the integrand as $\mathcal{L}[\eta^\alpha;\Psi_3]$, we see that is only dependent on position and normal displacement. Therefore, the Euler--Lagrange equations are simply $\mathcal{L}_{,\Psi_3}=0$, such that
\begin{equation}
(1+\nu) \left[Eh \left(h^2 \left[\kappa  R^2+\Psi_3 \right]+12 R^2 \Psi_3 \right)+12 (1-\nu) p R^4\right]=0.
\end{equation}
Solving for the normal displacement, we find that
\begin{equation}
\label{fund}
\Psi_3[p;\kappa]=-\frac{h^2}{12}\kappa-(1-\nu ) \frac{pR^2}{2 Eh}+\mathcal{O}\left(\frac{h^4}{R^4}\right)
\end{equation}
 It is clear that in the absence of either a natural curvature or a uniform pressure, equation~\ref{fund} reduces to the classical result from Hutchinson~\cite{Hutchinson1967} or the recent result by Pezzulla {\em et al.}~\cite{Pezzulla2018}, respectively. Since both stimuli act to change the normal displacement of the shell, there is a direct analogy between pressure and curvature~\cite{Pezzulla2018}, which is written
\begin{equation}
\label{analogy}
\kappa h = 6(1-\nu)\left(\frac{R}{h}\right)^2 \frac{p}{E}.
\end{equation}
This implies that a positive curvature stimulus $\kappa>0$ is analogous to a positive external pressure $p>0$, and both will result in the compression of the sphere. This foreshadows the results to come -- in the limit of small displacements and small strains, applying a positive $\kappa$ and positive $p$ simultaneously will act destabilize the shell, while oppositely signed stimuli may act to stabilize the shell. 

Turning to the question of stability, we need to evaluate the membrane and bending stress terms in equation~\ref{d2V}.  First, we will consider the uniform external pressure, which will only contribute to the membrane state of stress. The contravariant tensor of stress resultants from a uniform pressure is given by~\cite{koiter}
\begin{equation}
\label{Npab}
N^{\alpha\beta}\Big|_p=\sigma h \accentset{\circ}{a}^{\alpha\beta}=\frac{1}{2}pR\accentset{\circ}{a}^{\alpha\beta}.
\end{equation}
By using the pressure--curvature analogy given by equation~\ref{analogy}, we can write the contribution from pressure and curvature to the membrane stress resultants as
\begin{equation}
\label{Nab}
N^{\alpha\beta}=\frac{Eh}{2(1-\nu^2)}\left((1-\nu^2)\frac{p}{E}\frac{R}{h}+\frac{1+\nu}{6}\kappa h \frac{h}{R}\right)\accentset{\circ}{a}^{\alpha\beta},
\end{equation}
where the terms have been rearranged in accordance with equation~\ref{d2V}. Recalling equation~\ref{d2V}, and that $\xi_{\alpha\beta}$ can be determined by $\xi_{\alpha\beta}=\gamma_{\alpha\beta}-\varepsilon_{\alpha\beta}$ using equations~\ref{gammaab} and~\ref{ep}, we can write
\begin{equation}
\label{d2N}
\int_\omega N^{\alpha\beta}\xi_{\alpha\beta} \ \dd \omega=\frac{Eh}{2(1-\nu^2)}\int_\omega\left(\frac{1-\nu^2}{2}\frac{p}{E}\frac{R}{h}+\frac{1+\nu}{12}\kappa h \frac{h}{R}\right)\left[\left(\varepsilon^{\alpha\beta}-\omega^{\alpha\beta}\right)\left(\varepsilon_{\alpha\beta}-\omega_{\alpha\beta}\right)+\varphi_\alpha\varphi^\alpha\right] \ \dd \omega,
\end{equation}
where $\accentset{\circ}{a}^{\alpha\beta}$ from equation~\ref{Nab} was used to raise the indices of the nonlinear portion of the strain tensor. 

The remaining term to identify in equation~\ref{d2V} is the bending stress of the middle surface $M^{\alpha\beta}$. Since stability must be evaluated in the fundamental state, meaning the strain measure given by equation~\ref{varrhoNEP} is the relevant choice in determining $M^{\alpha\beta}$, such that
\begin{equation}
M^{\alpha\beta}=\frac{h^3}{12}\mathcal{A}^{\alpha\beta\lambda\mu}\overline{\varrho}_{\lambda\mu}.
\end{equation}
Because of equations~\ref{bbar} and~\ref{fund}, the bending strain becomes
\begin{equation}
\overline{\varrho}_{\alpha\beta}=\varrho_{\alpha\beta}-\kappa\accentset{\circ}{a}_{\alpha\beta}=\left(\frac{1}{12}\frac{h^2}{R^2}+\frac{1-\nu}{2}\frac{1}{\kappa h}\frac{p}{E}-1\right)\kappa\accentset{\circ}{a}_{\alpha\beta}\simeq -\kappa\accentset{\circ}{a}_{\alpha\beta},
\end{equation}
since the assumptions used in deriving the governing shell equations require that $p/E\ll1$ and $h^2/R^2\ll1$. Now, all that is needed is the trace of the bending strains, 
\begin{equation}
\label{Mab}
M^{\alpha\beta}\zeta_{\alpha\beta}=-\frac{Eh^3}{12(1-\nu)}\kappa\zeta_\alpha^\alpha,
\end{equation}
which is given by~\cite{Deserno2004}
\begin{equation}
\label{traceZeta}
\zeta_\alpha^\alpha=\frac{1}{R^2}\Psi^\alpha\nabla_\alpha\Psi_3-\frac{1}{R}|\nabla\Psi_3|^2+\frac{1}{R}\Psi_\alpha\triangle\Psi^\alpha-\nabla_\alpha\Psi_3\triangle\Psi^\alpha,
\end{equation}
where $|\nabla\Psi_3|$ denotes the absolute value. We note that the Laplace operator is defined by $\nabla^\alpha\nabla_\alpha( \cdot )=\triangle ( \cdot )$, and the bilaplacian or biharmonic operator is $\triangle\triangle ( \cdot ) = \triangle^2( \cdot )$.  We can now write the contribution of bending stresses to the second variation of the total potential energy as
\begin{equation}
\label{d2M}
\int_\omega M^{\alpha\beta}\zeta_{\alpha\beta} \ \dd \omega=-\frac{Eh}{2(1-\nu^2)}\int_\omega\frac{1+\nu}{6}\kappa h^2\left(\frac{1}{R^2}\Psi^\alpha\nabla_\alpha\Psi_3-\frac{1}{R}|\nabla\Psi_3|^2+\frac{1}{R}\Psi_\alpha\triangle\Psi^\alpha-\nabla_\alpha\Psi_3\triangle\Psi^\alpha\right) \ \dd \omega.
\end{equation}
Equations~\ref{d2N} and~\ref{d2M}, along with the definitions in equations~\ref{epS} and~\ref{rhoS}, provide a complete description of the stability of a complete spherical shell under uniform pressure and homogenous natural curvature, as defined by the second variation of the total potential energy given by equation~\ref{d2V}. 

\section{Reduction of Second Variation Energy}
\label{reduction}

In the following section, we will greatly simplify the strain energy in the fundamental state, transforming the equations into a form that can be readily analyzed by linear stability analysis. In particular, we will reformulate equation~\ref{d2V} in a coordinate--free form, relying on a combination of invariant operators. Through the application of the generalized Stoke's theorem on a closed surface, the resulting are significantly simplified. The reduction of the second variation of the total potential energy is a rather lengthy endeavor, and much of it can be found in the reference~\cite{koiter}. Here, we highlight the salient properties of the reduction, and show the contribution of natural curvature to the membrane and bending prestresses. 

Van der Neut was the first to recognize that it is useful to apply the Helmholtz decomposition to the tangential displacement field of a spherical shell~\cite{Neut1932}. This analysis was used extensively by Koiter~\cite{koiter, koiter1969}, in his subsequent analysis of the nonlinear buckling behavior of spherical shells under pressure, as well as by Niordson in studying the vibrations of complete spherical shells~\cite{Niordson1984}. This decomposition allows the representation the tangential displacement field as the sum of a solenoidal (divergence--free) and irrotational (curl--free) vector field, 
\begin{equation}
\label{splitMain}
\Psi_\alpha=\phi_{,\alpha}+\epsilon_{\alpha\lambda}\nabla^\lambda\psi,
\end{equation}
where $\phi$ and $\psi$ are a scalar potentials. The advantage of expressing $\Psi_\alpha$ in terms of $\phi$ and $\psi$ is that for a closed spherical shell we can obtain an equation in which $\psi$ appears uncoupled from both $\phi$ and $\Psi_3$, and that ultimately $\psi$ will vanish from the second variation of the total potential energy $\delta^2\mathcal{V}$. The equations that result from this reduction share many similarities with those found by Koiter~\cite{koiter, koiter1969}, such that so we will only highlight how the presence of a nonzero natural curvature alters the second variation of the total potential energy in the main text. The strain energy in stretching and bending the shell remain unchanged from Koiter's results~\cite{koiter}, leading to
\begin{subequations}
\begin{align}
\label{d2Um}
\delta^2\mathcal{U}_m&=\frac{Eh}{2(1-\nu^2)}\int(\triangle\phi)^2+\frac{1-\nu}{R^2}\phi\triangle\phi+\frac{2(1+\nu)}{R}\Psi_3\triangle\phi+\frac{2(1+\nu)}{R^2}\Psi_3^2\,\dd\omega, \\
\label{d2Ub}
\delta^2\mathcal{U}_b&=\frac{Eh}{2(1-\nu^2)}\int \frac{h^2}{12}(\triangle\Psi_3)^2\,\dd\omega.
\end{align}
\end{subequations}
As we will show, the presence of a natural curvature inducing stimulus will alter both the membrane and bending prestress. The contribution from the natural curvature for a spherical shell in the absence of a uniform external pressure was recently discussed by Pezzulla {\em et al.}~\cite{Pezzulla2018}, and we utilize some of that analysis here.

If we first consider the membrane prestress given by equation~\ref{Nab}, we can make use of the pressure--curvature analogy first presented by Pezzulla {\em et al.}~\cite{Pezzulla2018}, and given again here by equation~\ref{analogy}. With this analogy, we can directly use Koiter's results for the reduction of the membrane prestress~\cite{koiter1969, koiter} to rewrite that equation as a linear superposition of contributions from both pressure and natural curvature. From Koiter~\cite{koiter}, the membrane prestress was written
\begin{equation}
\label{d2PmKoiter}
\delta^2\widetilde{\mathcal{P}}_m=\frac{pR}{2}\int_\omega  (\triangle\phi)^2+4\frac{\Psi_3}{R}\triangle\phi+\Psi_3\triangle\Psi_3+\frac{2}{R^2}\Psi_3^2+(\triangle\psi)^2\,\dd\omega,
\end{equation}
where we note that the difference in sign in equation~\ref{d2PmKoiter} comes from the different convention used by Koiter regarding the orientation of the normal vector on the spherical shell. Koiter then only retained terms whose magnitudes are of the same order as those retained in moderate rotation shell theory. Specifically, to be in the elastic range we must ensure that $\sigma/E \ll 1$, which introduces a relative error of $\mathcal{O}(\sigma/E) \sim 1$. Terms containing $h^2/R^2$ can also be neglected, as this introduces a a relative error of $\mathcal{O}(h^2/R^2)$, which is smaller than the error in the underlying shell theory. In addition, it is demonstrated in~\cite{koiter} that $\triangle\psi=0$. Thus, Koiter neglected all terms in equation~\ref{d2PmKoiter} except~$\Psi_3\triangle\Psi_3$, by comparing them with similar ones in the elastic energy and showing that they are smaller by at least a factor~$h/R$. We shall do the same here, and using equation~\ref{analogy}, we write the second variation of the membrane prestress as
\begin{equation}
\label{d2Pm}
\delta^2\mathcal{P}_m=\frac{Eh}{2(1-\nu^2)}\int \left[\frac{1+\nu}{6}\frac{h}{R}\kappa h\Psi_3\triangle\Psi_3+(1-\nu^2)\frac{R}{h}\frac{p}{E}\Psi_3\triangle\Psi_3 \right]\dd\omega.
\end{equation}

Turning our attention to the bending prestress, we have to apply the decomposition of $\Psi_\alpha$ to equation~\ref{d2M}, which is helpful to do term--by--term. Let us start with the first term in equation~\eqref{d2M}, and rewrite it by using the chain rule
\begin{equation}
\begin{aligned}
\frac{1}{R^2}\int\Psi^\alpha\nabla_\alpha\Psi_3\, \dd\omega=\frac{1}{R^2}\int\nabla_\alpha(\Psi^\alpha\Psi_3)-\triangle\phi\Psi_3\, \dd\omega=-\frac{1}{R^2}\int\triangle\phi\Psi_3\,\dd\omega,
\end{aligned}
\end{equation}
where we utilized the symmetry of~$\nabla_{\alpha\beta}\psi$ to write~$\nabla_\alpha\Psi^\alpha=\triangle\phi$, and then applied the generalized Stokes theorem on the term containing $\nabla_\alpha(\Psi^\alpha\Psi_3)$, which of course vanishes on a closed spherical shell. Application of the generalized Stokes theorem is helpful once again for the second term in equation~\eqref{d2M}, allowing us to write
\begin{equation}
\begin{aligned}
-\frac{1}{R}\int|\nabla\Psi_3|^2\,\dd\omega=-\frac{1}{R}\int\nabla_\alpha(\nabla^\alpha\Psi_3)-\Psi_3\triangle\Psi_3\,\dd\omega=\frac{1}{R}\int\Psi_3\triangle\Psi_3\,\dd\omega\,,
\end{aligned}
\end{equation}
where again the divergence term disappears by use of the generalized Stokes theorem on a closed surface. For the third term in equation~\eqref{d2M} we have
\begin{equation}\label{psiapsia}
\begin{aligned}
\frac{1}{R}\int\Psi_\alpha\triangle\Psi^\alpha\,\dd\omega=\frac{1}{R}\int\nabla_\alpha\phi\nabla^{\alpha\beta}_{\cdot\cdot\beta}\phi+\varepsilon_{\alpha\lambda}\nabla^\lambda\psi\nabla^{\alpha\beta}_{\cdot\cdot\beta}\phi+\nabla_\alpha\phi\varepsilon^\alpha_\gamma\nabla^{\gamma\cdot\beta}_{\cdot\beta}\psi+\varepsilon_{\alpha\gamma}\nabla^\gamma\psi\varepsilon^\alpha_\lambda\nabla^{\lambda\beta}_{\cdot\cdot\beta}\psi\,\dd\omega\,.
\end{aligned}
\end{equation}
As we noted earlier, the advantage of this vector field decomposition is that all terms coupling $\psi$ and $\phi$ and $\psi$ and $\Psi_3$ vanish. To see how these terms disappear, we point the reader to Appendix A, or to the derivation of similar terms found in~\cite{koiter}. As a consequence of this relation, the second and third in~\eqref{psiapsia} terms are zero. For the first term, we can change the order of covariant differentiation, taking care to account for the inequality of the covariant derivatives, {\em i.e.} $\nabla^{\alpha\beta}_{..\beta}\phi \neq \nabla^{\alpha.\beta}_{.\beta}\phi$. The difference in the covariant derivatives is accounted for through the Riemann--Christoffel curvature tensor~\cite{Niordson1985}, which is related to the Gaussian curvature $\mathcal{K}=R^{-2}$ in the fundamental state by 
\begin{equation}
\label{riemann}
\mathcal{R}^\kappa_{.\alpha\beta\lambda}=\mathcal{K}\epsilon^{\kappa}_{.\alpha}\epsilon_{\beta\lambda},
\end{equation}
such that
\begin{equation}
\frac{1}{R}\int\nabla_\alpha\phi\nabla^{\alpha\beta}_{\cdot\cdot\beta}\phi\,\dd\omega=-\frac{1}{R}\int_\omega(\triangle\phi)^2+\frac{\triangle\phi}{R^2}\phi \ \dd\omega,
\end{equation}
where we utilized the relation between the Levi--Civita symbol and the metric tensor, {\em i.e.} $\epsilon^{\kappa\alpha}\epsilon^\beta_{. \alpha}=\accentset{\circ}{a}^{\kappa\beta}$, used the chain rule, and then applied the generalized Stoke's theorem on each of the terms. By analogy, the fourth term can be rewritten as
\begin{equation}
\frac{1}{R}\int\varepsilon_{\alpha\gamma}\nabla^\gamma\psi\varepsilon^\alpha_\lambda\nabla^{\lambda\beta}_{\cdot\cdot\beta}\psi\,\dd\omega=-\frac{1}{R}\int_\omega(\triangle\psi)^2+\frac{\triangle\psi}{R^2}\psi \ \dd\omega = 0,
\end{equation}
which vanishes due to the identity $\triangle\psi=0$~\cite{koiter}. Finally, for the fourth term in equation~\eqref{d2M} we find
\begin{equation}
\begin{aligned}
-\int\nabla_\alpha\Psi_3\triangle\Psi^\alpha\,\dd\omega=\int\nabla_{\alpha\beta}\Psi_3\nabla^{\alpha\beta}\phi\,\dd\omega=\int\triangle\phi\Bigl(\triangle\Psi_3+\frac{\Psi_3}{R^2}\Bigr)\,\dd\omega\,.
\end{aligned}
\end{equation}
Using these simplifications, the contribution of the bending moments in the fundamental state becomes
\begin{equation}\label{moments}
\begin{aligned}
\delta^2\mathcal{P}_b=-\frac{Eh^2}{12(1-\nu)}\kappa h\int\frac{1}{R}\Psi_3\triangle\Psi_3+\triangle\phi\triangle\Psi_3-\frac{(\triangle\phi)^2}{R}-\frac{\triangle\phi}{R^3}\phi\,\dd\omega\,.
\end{aligned}
\end{equation}
Now, since the term proportional to~$(\triangle\phi)^2$ in equation~\eqref{moments} has a smaller pre-factor than it does in equation~\eqref{d2PmKoiter}, which Koiter neglected, it can be neglected here as well. Moreover, the term coupling $\phi$ and $\triangle \phi$ has a much smaller prefactor than the term in equation~\ref{d2Um}, {\em i.e.} $(h/R)^3 \ll h/R$, and it too can be neglected. Finally, the contribution of the bending moments in the fundamental state can be reduced to
\begin{equation}
\label{d2Pb}
\delta^2\mathcal{P}_b=\frac{Eh}{2(1-\nu^2)}\int -\frac{1+\nu}{6}\kappa h^2\left[\frac{1}{R}\Psi_3\triangle\Psi_3+\triangle\phi\triangle\Psi_3\right]\,\dd\omega.
\end{equation}

Additional subtleties regarding the terms retained in the equations~\ref{d2Um} and~\ref{d2Ub} below are discussed in detail in~\cite{koiter}, and these results are unaffected by the natural curvature stimulus considered in this paper. Finally, the second variation of the total potential energy in the fundamental state consists of the four functionals given by equations~\ref{d2Um},~\ref{d2Ub},~\ref{d2Pm}, and~\ref{d2Pb}, such that we may write
\begin{equation}
\label{d2Vsimple}
\delta^2\mathcal{V}[\vett{\Psi};\kappa, p]=\delta^2\mathcal{U}_m[\vett{\Psi}]+\delta^2\mathcal{U}_b[\vett{\Psi}]+\delta^2\mathcal{P}_m[\vett{\Psi};\kappa, p]+\delta^2\mathcal{P}_b[\vett{\Psi};\kappa].
\end{equation}
The distinction between these equations and those found by Koiter~\cite{koiter1969, koiter} are evident by the presence of $\kappa$ and $p$ altering the membrane prestress, and the contribution of the natural curvature $\kappa$ to the bending prestress, a term typically neglected.

\section{Linear Stability Analysis -- Critical Buckling Pressure}
\label{stabilityAnalysis}

The necessary condition for a minimum of $\delta^2\mathcal{V}[\vett{\Psi};\kappa, p]=\delta^2\mathcal{U}_m+\delta^2\mathcal{U}_b+\delta^2\mathcal{P}_m+\delta^2\mathcal{P}_b$ is that its first variation with respect to $\phi$ and $\Psi_3$ vanishes. For example, let us consider the first term of equation~\ref{d2Um}. The first variation of the term $\int(\triangle \phi)^2 \ \dd \omega$ is found by application of the chain rule, followed by successive integration by parts, yielding the result
\begin{equation}
\begin{split}
2\int_\omega\triangle\phi\triangle\delta\phi \ \dd \omega&=2\int_\omega\triangle\phi\nabla_\alpha\nabla^\alpha\delta\phi \ \dd \omega=2\left[\triangle\phi \nabla^\alpha\delta\phi \right] \Big|_\Gamma - 2\int_\omega\nabla_\alpha(\triangle \phi)\nabla^\alpha \delta\phi \ \dd\omega \\
&=-2\left[\nabla^\alpha(\triangle\phi)\delta\phi \right] \Big|_\Gamma + 2\int_\omega\nabla^\alpha\nabla_\alpha(\triangle\phi) \delta\phi \ \dd \omega =2\int_\omega \triangle^2 \phi \ \delta\phi \ \dd \omega.
\end{split}
\end{equation}
The integration by parts results in boundary terms to be evaluated on the boundary $\Gamma$, which of course is a boundary of zero length for a closed spherical shell, and so these terms disappear. This analysis can be applied term by term in equations~\ref{d2Um},~\ref{d2Ub},~\ref{d2Pm}, and~\ref{d2Pb} --  a tedious, but straightforward exercise that is not shown here for the sake of brevity. Collecting terms with $\delta \phi$ and $\delta \Psi_3$ separately, we arrive at the following two equations that must be satisfied to be in neutral equilibrium
\begin{subequations}
\begin{align}
\frac{h^2}{12}\triangle^2\Psi_3+\frac{1+\nu}{R}\triangle \phi+\frac{2(1+\nu)}{R^2}\Psi_3+\frac{1-\nu^2}{2}\frac{R}{h}\frac{p}{E}\triangle\Psi_3-\frac{1+\nu}{12}\frac{h}{R}\kappa h\triangle\Psi_3-\frac{1+\nu}{12}\kappa h^2\triangle^2\phi&=0, \\
\triangle^2 \phi+\frac{1-\nu}{R^2}\triangle\phi+\frac{1+\nu}{R}\triangle\Psi_3-\frac{1+\nu}{12}\kappa h^2\triangle^2\Psi_3&=0.
\end{align}
\end{subequations}
Since the displacement vectors exist on the surface of a complete spherical shell, our analysis will be aided by expanding $\phi$ and $\Psi_3$ in a series of spherical surface harmonics
\begin{subequations}
\begin{align}
\label{phiSurf}
\phi(\eta^\alpha)&=R\sum_{n=0}^{\infty}D_nS_n(\eta^\alpha), \\
\label{psiSurf}
\Psi_3(\eta^\alpha)&=\sum_{n=0}^\infty C_nS_n(\eta^\alpha), 
\end{align}
\end{subequations}
where $S_n(\eta^\alpha)$ is a spherical surface harmonic of degree $n$, described by the differential equation
\begin{equation}
\label{Sn}
\Delta S_n(\eta^\alpha) = -\frac{1}{R^2}n(n+1)S_n(\eta^\alpha).
\end{equation}
Expanding the equations of neutral equilibrium in a series of spherical harmonics leads to the following eigenvalue problem for the critical buckling pressure in terms of the natural curvature, material properties, and shell geometry
\begin{equation}\label{system}
\begin{aligned}
-(1+\nu)\Bigl[1+\frac{\kappa h}{12}\frac{h}{R} x\Bigr]C_n+[x-(1-\nu)]D_n&=0\,,\\
\Bigl[\frac{1}{12}\Bigl(\frac{h}{R}\Bigr)^2x^2+\frac{1+\nu}{12}\frac{h}{R}\kappa h x -\frac{1-\nu^2}{2}\frac{R}{h}\frac{p}{E}+2(1+\nu)\Bigr]C_n-(1+\nu)\Bigl[x+\frac{\kappa h}{12}\frac{h}{R} x^2\Bigr]D_n&=0\,,
\end{aligned}
\end{equation}
\begin{figure}[t]
\begin{center}
\includegraphics[width=1\columnwidth]{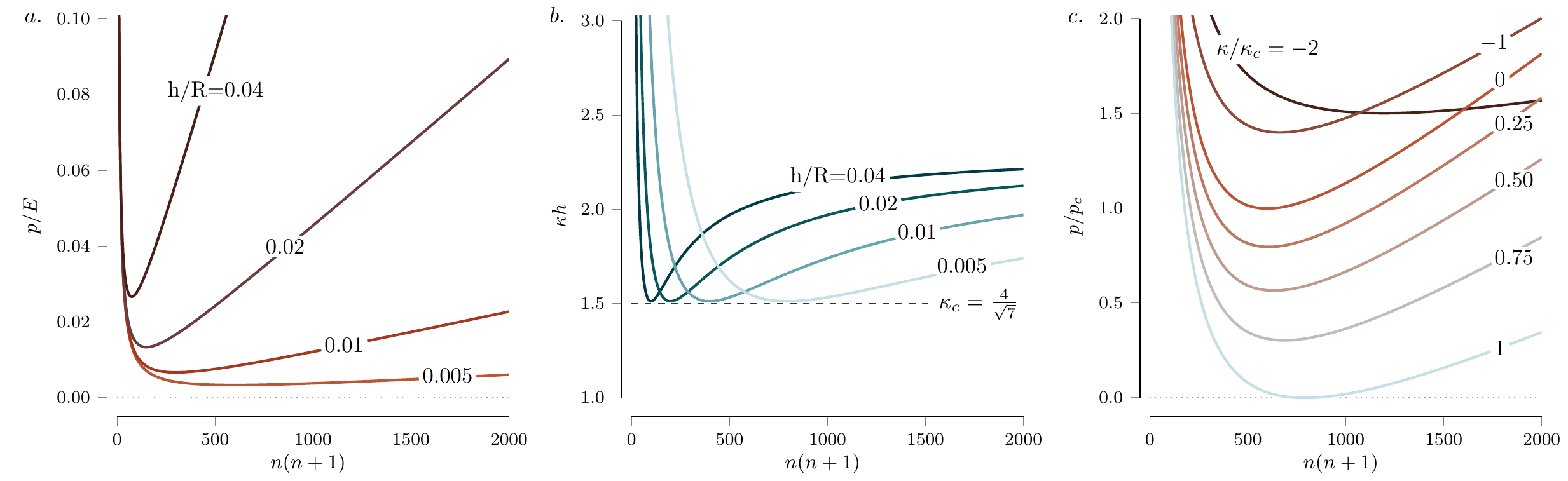}
\vspace{-6mm}
\caption{\label{eigenmodes} $a.$ A plot of externally applied pressure $p$ normalized by the Young's modulus $E$ as a function of eigenmodes written as $n(n+1)$ for several values of $h/R$. This plot is the classical result in the absence of natural curvature $\kappa$. $b.$ A plot of natural curvature $\kappa$ normalized by $h$ as a function of eigenmodes written as $n(n+1)$ for several values of $h/R$. This plot is in the absence of external pressure, and recreates the results from reference~\cite{Pezzulla2018}. $c.$ A plot illustrating the coupling of pressure and curvature. Here, the pressure is normalized by the classical buckling pressure given by equation~\ref{pc} and is plotted against eigenmodes written as $n(n+1)$ for several values of natural curvature normalized by the critical buckling curvature $\kappa_c=4/\sqrt{7}$. These curves are plotted from equation~\ref{pEx}.}
\end{center}
\end{figure}
where~$x=n(n+1)$. The result of this eigenvalue problem is
\begin{equation}
\label{pEfull}
\frac{p}{E}=\frac{1}{72}\frac{h}{R}\left(\left[\frac{6x}{1+\nu}+\frac{2x(3-\kappa^2h^2)}{1-\nu}\right]\left(\frac{h}{R}\right)^2+\frac{(2-x)(12+\kappa h x \frac{h}{R})^2}{x(1-x-\nu)}-\frac{36 \kappa h}{1-\nu}\frac{h}{R}\right)
\end{equation}
As is true for pressure buckling of shells in the absence of natural curvature, values of $n$ of $\mathcal{O}(1)$ would lead to stresses in the shell of $p/E = \mathcal{O}(1)$, which would take the shell outside of the elastic range. To be consistent with the small strain approximation used in developing the shell equations, staying in the elastic range requires that $p/E \ll 1$, which will only occur is $n \gg 1$, and therefore $x \gg 1$. We can therefore simplify equation~\ref{pEfull} to yield
\begin{equation}
\label{pEx}
\frac{p}{E}=\frac{1}{72}\frac{h}{R}\left[\frac{\left(12+\frac{h}{R} \kappa h n\right)^2}{n}-\frac{2\left(18 \kappa h + \frac{h}{R}n[\kappa^2 h^2-3]\right)}{1-\nu}\frac{h}{R}+\frac{6n}{1+\nu}\left(\frac{h}{R}\right)^2\right]
\end{equation}
which gives an equation that is quadratic in $x$. To obtain the critical value of $p/E$ we now minimize equation~\ref{pEx} with respect to $x$, which yields
\begin{equation}
\label{x}
n(n+1)=12\left(\frac{R}{h}\right)\frac{\sqrt{1-\nu^2}}{\sqrt{12-\kappa^2h^2 (1+\nu)^2}}.
\end{equation}
Finally, the critical buckling pressure for a shell with a natural curvature stimulus is found by inserting the eigenmode from equation~\ref{x} into equation~\ref{pEx} and rearranging the result to find
\begin{equation}
\label{pEkhc}
p_\kappa=2 E\left(\frac{h}{R}\right)^2 \left[\sqrt{\frac{1}{3 \left(1-\nu ^2\right)}-\frac{\kappa^2 h^2 (1+\nu)^2}{36 \left(1-\nu ^2\right)}}-\frac{\kappa h(1+2 \nu)}{12 (1-\nu )}\right].
\end{equation}
It is immediately clear that in the absence of a natural curvature stimulus, {\em i.e.} $\kappa=0$, equation~\ref{pEkhc} reduces to the classical buckling pressure of a spherical shell obtained by Zoelly~\cite{Zoelly1915}
\begin{equation}
\label{pc}
p_c = \frac{2E}{\sqrt{3(1-\nu^2)}}\left(\frac{h}{R}\right)^2
\end{equation}
\begin{figure}[t]
\begin{center}
\includegraphics[width=1\columnwidth]{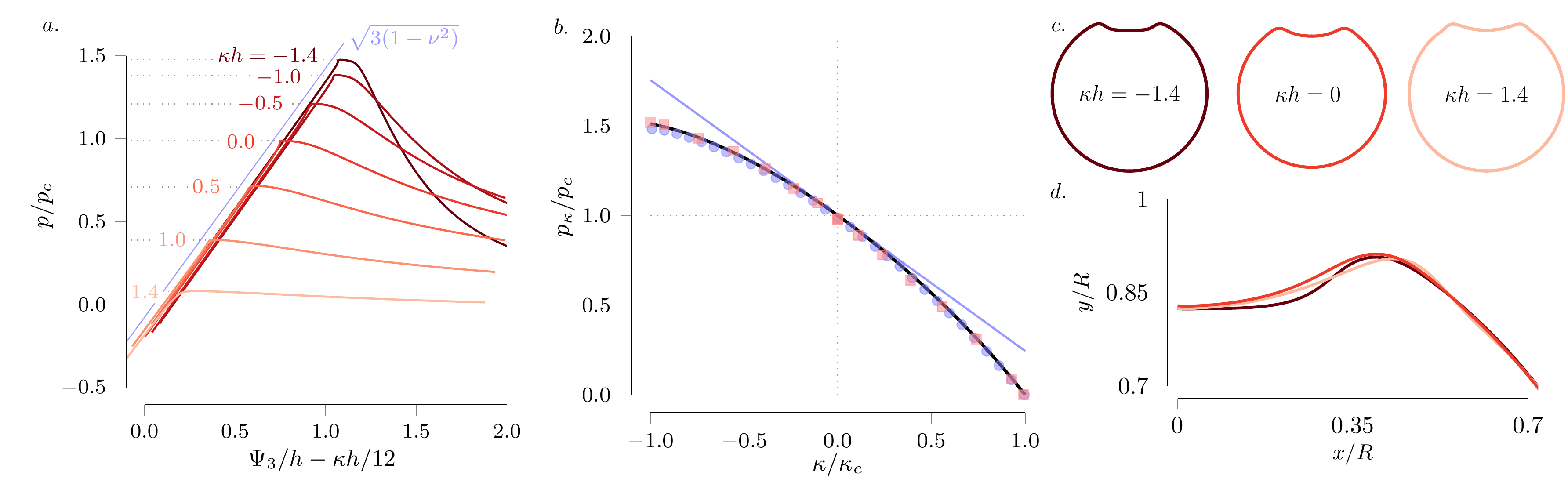}
\vspace{-6mm}
\caption{\label{pk} $a.$ Plots of the pressure normalized by $p_c$ as a function of the normal displacement of the shell at its apex. The critical pressure $pc$ is given by equation~\ref{pc} and the normal displacement along the $x$--axis is given according to equation~\ref{fund}. As normalized, the slope of the fundamental state is given by $\sqrt{3(1-\nu^2)}$. We not that the slope of the post buckling curve becomes very steep for large, negative values of $\kappa h$, and is nearly flat, but still decreasing, for large, positive values of $\kappa h$. $b.$ The theoretical curve from equation~\ref{pEkhc}, along with its linearized form given by equation~\ref{pEkhcLIN}. The $x$--axis is normalized by the critical buckling curvature, given by equation~\ref{kc}. Numerical results from the 1D axisymmetric model (\protect\tikz \protect\draw[blue!40,fill=blue!40] (0,0) circle (.55ex);) and the 2D model (\protect\tikz \protect\draw[red!50,fill=red!50] (0,0) rectangle (0.15,0.15);) coincide directly with the analytical prediction. $c.$ Postbuckling shapes from the 1D model at $\Psi_3/h=20$ for three different values of $\kappa h$. $d.$ Profiles of the the postbuckling shapes that have been magnified near the shell's apex.}
\end{center}
\end{figure}
With a little bit of algebra, it can also be shown that when $p=0$, we recover the critical buckling curvature obtained by Pezzulla~{\em et al.}~\cite{Pezzulla2018}
\begin{equation}
\label{kc}
\kappa_c = \frac{4}{h}\sqrt{3\frac{1-\nu}{(1+\nu)(5+4\nu)}}.
\end{equation}
Additional insight into equation~\ref{pEkhc} can be gleaned by considering the shell incompressible, {\em e.g.} $\nu=1/2$, and linearizing the result.
\begin{equation}
\label{pEkhcLIN}
\frac{p_\kappa}{p_c} \approx 1 - \frac{\kappa}{2\kappa_c}
\end{equation}
This result is instructive for three reasons. First, we immediately see that a positive natural curvature ({\em e.g.} a bilayer shell with an outer layer that is shrinking) acts in concert with the positive externally applied pressure to destabilize the shell. Second, as noted in~\cite{Pezzulla2018}, the curvature strains needed to buckle a complete spherical shell are not small. Indeed, curvature strains for {\em Volvox} eversion are often cited to be between $1.5  < \kappa h < 3$~\cite{Hohn2011, Hohn2015}, and while these shells are often analyzed using a Helfrich model, it is reasonable to expect that finite strain constitutive shell models may be needed to fully quantify curvature induce buckling and its analogous limit point instabilities with full spherical shells. We expect that the linearized result given by equation~\ref{pEkhcLIN} will remain valid in the small strain range. Third, and rather intriguingly, this result  suggests that when a negative natural curvature stimulus is applied it should strengthen the shell against pressure buckling. To check the analytical prediction, we performed numerical simulations with both a 1D axisymmetric shell model (\ref{1D}) and a 2D shell model (\ref{2D}). In figure~\ref{pk}a, we plot numerical results from the 1D model of $p/p_c$ as a function of the normalized normal displacement of the shell (see: equation~\ref{fund}), for a wide range of $\kappa h$ values. In these simulations, was taken to be $\nu=1/2$, and $R/h$ was varied slightly, but always remained above $R/h=90$ (\ref{1D}). We see in figure~\ref{pk}b that the critical buckling pressure from both the 1D and 2D models coincide directly with the theoretical prediction from equation~\ref{pEkhc}. Figure~\ref{pk}b represents the key finding of this work, {\em i.e.} that a negative natural curvature stimulus should strengthen a shell against pressure buckling. The slope of the posbuckling curve appears to be strongly dependent on the value of $\kappa h$, indicating that while negative values of $\kappa h$ appear to increase the pressure required to buckle a shell, they simultaneously may be making the shell significantly more sensitive to imperfections. Indeed, the post buckling shape of the shells is appears qualitatively different as $\kappa h$ decreases ($\Psi_3/h=20$, Fig.~\ref{pk}c). From the 1D model, we see that for $\kappa h =1.4$ ($\kappa/\kappa_c =0.926$) the dimple at the apex has a curvature that slowly varies along the arc length, while for $\kappa h =-1.4$ ($\kappa/\kappa_c =-0.926$), the apex of the shell is nearly flat, and there is a high curvature at the inflection point and dimple ridge (Fig.~\ref{pk}d). Both postbuckling shapes are qualitatively different than the classical dimple shape of a shell under uniform pressure. 

\section{Conclusions}

In summary, we have presented an analytical study of the buckling of a complete spherical shell under the combined loading of external pressure and a non--mechanical stimulus that induces a change in the shells natural curvature. We extended the classical results found by Koiter~\cite{koiter1969, koiter} to include the affects of the bending prestress, and we showed that an evolving natural curvature can significantly increase or decrease the pressure required to induce buckling. The analytical model was verified numerically using both a 1D axisymmetric model and a 2D model, and the theory and numerics suggest that a negative natural curvature could produce a {\em knock--up factor} in the buckling of spherical shells and caps. This could be achieved experimentally by preparing thin shells out bimetals and studying shell buckling under different temperatures. The numerical results motivate further study into the imperfection sensitivity and postbuckling behavior of shells under a combined pressure and curvature loading, and in particular we would advocate for following the work established by Hutchinson in this regard~\cite{Hutchinson1967, Hutchinson2016}. In addition, we note that in the absence of pressure, or in the presence of pressures $p/p_c < 1$, the model suggests that curvature strains on the order of $\kappa h \sim \mathcal{O}(1)$ are needed to induce buckling. So, this implies that while strains that measure the stretching of the middle surface are small, as is consistent with the formulation of shell theory, the bending strains may in fact be finite. This implies that we might expect material nonlinearity at large positive and negative values of $\kappa/\kappa_c$. These implications motivate further work to examine the buckling of a spherical shell with a neo--Hookean or Mooney--Rivlin constitutive model to see if there is any stiffening in the shell at large values of $\kappa h$.

\section{Acknowledgements}
DPH, JHL, and HSP gratefully acknowledge the financial support from NSF through CMMI-1824882. DPH and MP are grateful for the discussion we shared with John Hutchinson during the early stages of this manuscript.

\setcounter{section}{0}
\renewcommand{\thesection}{Appendix \Alph{section}}
\numberwithin{equation}{section}
\renewcommand\theequation{\Alph{section}.\arabic{equation}}

\section{1D model}
\label{1D}
To test the analytical results found in this paper, we implement an 1D, axisymmetric shell model in COMSOL Multiphysics, using its nonlinear, Weak Form PDE solver. To follow the shell's deformation through the instability, an arc--length method is implemented through the inclusion of a Domain Point Probe at the apex of the shell that moves vertically. A stationary sweep of the vertical displacement was performed in increments of at least $-h/3000$. Example parameters are given in table~\ref{params}. To construct the equations to be used within COMSOL, we begin by parameterizing the shell. The surface of a sphere of radius $R$ in spherical coordinates denoted by its inclination $\phi$ and azimuth $\psi$ is parameterized by 
\begin{equation}
\accentset{\circ}{\vett{r}}(\phi,\psi)=(R \cos \psi \sin \phi ,R \sin \psi  \sin \phi ,R \cos \phi ), 
\end{equation}
and we will look for deformed shapes with rotational symmetry parameterized by 
\begin{equation}
\vett{r}(\phi, \psi)=(f(\phi) \cos \psi, f(\phi) \cos \psi, g(\phi) ).
\end{equation}
In the absence of pressure and a curvature stimulus, {\em i.e.} $p=\kappa=0$, $\vett{r} \rightarrow \accentset{\circ}{\vett{r}}$ when $f(\phi)=R \sin \phi$ and $g(\phi)=R \cos \phi$.
The unit vectors on the surface $\accentset{\circ}{\vett{a}}_\phi$ and $\accentset{\circ}{\vett{a}}_\psi$, and the outward normal vector $\vett{n}$ are given by equations~\ref{eq-aVec} and~\ref{eq-n}, respectively. From here the metric tensor can be derived as $\accentset{\circ}{\vett{a}}=\accentset{\circ}{\vett{a}}_\alpha\cdot\accentset{\circ}{\vett{a}}_\beta$, and the curvature tensor is derived as $\accentset{\circ}{\vett{b}} = \accentset{\circ}{\vett{a}}_{\alpha,\beta} \cdot \accentset{\circ}{\vett{n}}$, such that the metric and curvature tensors of the reference and deformed spherical shells are
\begin{equation}
\begin{aligned}[c]
&\accentset{\circ}{\vett{a}}=  
\begin{pmatrix}
R^2 &0\\
0& R^2 \sin^2 \phi\\
\end{pmatrix},\\
&\accentset{\circ}{\vett{b}}=
\begin{pmatrix}
-R&0\\
0&-R \sin^2{\phi}\\
\end{pmatrix},
\end{aligned}
\qquad\qquad
\begin{aligned}[c]
&\vett{a}=
\begin{pmatrix}
f_{,\phi}^2+g_{,\phi}^2 &0\\
0&f^2\\
\end{pmatrix},\\
&\vett{b}=\frac{1}{\sqrt{f_{,\phi}^2+g_{,\phi}^2}}
\begin{pmatrix}
g_{,\phi\phi}f_{,\phi}-f_{,\phi\phi}g_{,\phi}&0\\
0&fg_{,\phi}\\
\end{pmatrix}.
\end{aligned}
\end{equation}
\begin{table}
\begin{tabular*}{0.94\columnwidth}{@{\extracolsep{\fill}}lll} 
\toprule
	{\bf Name}	&	{\bf Expression}		&	{\bf Description}				\\	
\toprule
	\tt{thick} 	&	1 			&	Shell thickness 				\\ 
	\tt{theta} 	&	\tt{pi}			&	Angle from pole to pole 		\\
	\tt{ni}		&	1/2			&	Poisson's ratio				\\
	\tt{n}	&	20			&	Mode number at buckling		\\
	\tt{kappa}	&	1.4			&	Natural curvature 			\\
	\tt{R}		&	\tt{n$\ast$(n+1)$\ast$sqrt(12-(kappa$\ast$h$\ast$(1+ni))\^{}2)/(12$\ast$sqrt(1-ni\^{}2))}		&	Shell Radius (calculated from eq.~\ref{x})	\\
	\tt{Rh}	&	R/thick		& 	Radius to thickness ratio		\\
	\tt{V}		&	4/3$\ast$pi$\ast$R\^{}3	&	Initial enclosed volume \\	
	\tt{Amp}	&	12			&	Amplitude of eigenmode imperfections \\
	\tt{c}		&	sqrt(3$\ast$(1-ni\^{}2))	&	Constant used in pb	\\
	\tt{pb}	&	2/(c$\ast$Rh\^{}2)	&	Critical buckling pressure (Zoelly)	\\
	\tt{Lambdao}	&	1		& 	Stimuli-induced stretching		\\
	\tt{p}		&	0			&	Initial pressure				\\
	\tt{disp}	&	0			&	Initial displacement at north pole \\
	
\bottomrule
\end{tabular*}
\caption{Example parameters used in COMSOL model. The mode number {\tt nn} and imperfection amplitude {\tt Amp} are chosen arbitrarily, and can be changed as desired.\label{params}}
\end{table}
With these definitions, the dimensionless stretching and bending energies can be written as
\begin{subequations}
\begin{align}
\overline{\mathcal{U}}_s&=\int\left((1-\nu)\left[\accentset{\circ}{a}_{11}^{-2}(a_{11}-\accentset{\circ}{a}_{11})^2+\accentset{\circ}{a}_{22}^{-2}(a_{22}-\accentset{\circ}{a}_{22})^2\right]+\nu\left[\accentset{\circ}{a}_{11}^{-1}a_{11}+\accentset{\circ}{a}_{22}^{-1}a_{22}-2\right]^2\right)\sqrt{\accentset{\circ}{a}_{11}\accentset{\circ}{a}_{22}}\ \dd A \\
\overline{\mathcal{U}}_b&=\Lambda^2\frac{h^2}{3}\int\left((1-\nu)\left[\accentset{\circ}{a}_{11}^{-2}(b_{11}-\accentset{\circ}{b}_{11})^2+\accentset{\circ}{a}_{22}^{-2}(b_{22}-\accentset{\circ}{b}_{22})^2\right]+\nu\left[\accentset{\circ}{a}_{11}^{-1}b_{11}+\accentset{\circ}{a}_{22}^{-1}b_{22}-2\accentset{\circ}{H}\right]^2\right)\sqrt{\accentset{\circ}{a}_{11}\accentset{\circ}{a}_{22}} \ \dd A,\\
\end{align}
\end{subequations}
where $\accentset{\circ}{H}$ is the mean curvature as defined by $2\accentset{\circ}{H}=\accentset{\circ}{b}_\alpha^\alpha$, the energies are integrated over the area of the sphere, and the energies have been nondimensionalized as
\begin{equation}
\label{nondimU}
\overline{\mathcal{U}}=\frac{8(1-\nu^2)}{Eh}\mathcal{U}.
\end{equation}
Likewise, the dimensionless potential of the dead pressure and natural curvature are nondimensionalized in the same manner as equation~\ref{nondimU}, and are given by
\begin{subequations}
\begin{align}
\overline{\mathcal{P}}_p&=\frac{8(1-\nu^2)}{h}\frac{p}{E}\int\Psi_3\sqrt{\accentset{\circ}{a}_{11}\accentset{\circ}{a}_{22}}\ \dd A,\\
\overline{\mathcal{P}}_\kappa&=-\frac{2(1+\nu)}{3}h^2\kappa\int\accentset{\circ}{a}_{11}^{-1}(\accentset{\circ}{b}_{11}-\accentset{\circ}{b}_{11})+\accentset{\circ}{a}_{22}^{-1}(b_{22}-\accentset{\circ}{b}_{22})\sqrt{\accentset{\circ}{a}_{11}\accentset{\circ}{a}_{22}}\ \dd A,
\end{align}
\end{subequations}
where since $E$ is not a given parameter in table~\ref{params}, the model instead calculates the dimensionless pressure $\overline{p}=p/E$. The normal deflection $\Psi_3$ is calculated from the unknown components of the deformed shell, $f_1$ and $g_1$, {\em i.e.}
\begin{equation}
\Psi_3=g_1\cos\phi+f_1\sin \phi.
\end{equation}
The shell is seeded with imperfections using a superposition of several eigenmodes of the buckling pattern, such that the initial shell shape is given by
\begin{subequations}
\begin{align}
f_0&=R\left(\sin \phi+\frac{h}{\beta R}\sin(n\phi)\sin\phi+\frac{h}{\beta R}\sin(2n\phi)\sin\phi+\frac{h}{2 \beta R}\sin(3n\phi)\sin\phi\right)\\
g_0&=R\left(\cos \phi+\frac{h}{\beta R}\sin(n\phi)\cos\phi+\frac{h}{\beta R}\sin(2n\phi)\cos\phi+\frac{h}{2 \beta R}\sin(3n\phi)\cos\phi\right),
\end{align}
\end{subequations}
where $\beta$ is the amplitude {\tt Amp} and $n$ is the mode number {\tt nn} found in tabe~\ref{params}, and the deformed shell shape is found using
\begin{subequations}
\begin{align}
f(\phi)&=f_0+f_1,  \\
g(\phi)&=g_0+g_1.\\
\end{align}
\end{subequations}
The two components of the shell shape are only a function of the unknown variable $\phi$, so $\phi$ and its first and second partial derivatives compose the weak form of the total potential energy that is minimized with COMSOL. To close the problem, we specify no boundary conditions such that there is no horizontal displacement at the north pole, no horizontal or vertical displacement at the south pole, and no slope in either direction at each pole.

\section{2D Model}
\label{2D}
The 2D simulations were carried out using an isogeometric analysis (IGA)-based discretization of Kirchhoff-Love shell theory under large deformations for the sphere in which the thickness was 2 mm and the radius was 5 cm.  The IGA is well-suited to modeling shell problems as it naturally incorporates higher order field approximations, which satisfies the C1 continuity requirement that arises due to the second derivatives that are present in the Kirchhoff-Love shell formulation.  Additionally, non-uniform rational B-splines (NURBS), which are used as the shape function in the IGA, can be used for curved-shell problems since they exactly model all conic sections such as circles, spheres, and ellipsoids~\cite{Cottrell2009}.  Details of the Kirchhoff-Love shell formulation under large deformations including the IGA computational formulation are omitted here, but can be found in~\cite{Sauer2017,Duong2017}.  The weak form is given with the admissible variation $\delta\bm{r} \in \mathcal{V}$ by
\begin{equation}
    \label{eq:weakformfull}
    G_{in} + G_{int} - G_{ext} = 0 \quad \forall\delta\bm{r} \in \mathcal{V}
\end{equation}

\begin{equation}
    \label{eq:weakformin}
    G_{in}=\int_{S_{0}} \delta\bm{r}\cdot\rho_{0}\dot{\bm{v}} dA
\end{equation}

\begin{equation}
    \label{eq:weakformint}
    G_{int}=
    \int_{S_{0}} \frac{1}{2}\delta a_{\alpha\beta}\tau^{\alpha\beta} dA 
    + \int_{S_{0}} \delta b_{\alpha\beta}M_{0}^{\alpha\beta} dA
\end{equation}

\begin{equation}
    \label{eq:weakformext}
    G_{ext}=
    \int_{S_{0}} \delta\bm{r}\cdot\bm{f} da
    + \int_{\partial_{t}S} \delta\bm{r}\cdot\bm{t} ds
    + \int_{\partial_{m}S} \delta\bm{n}\cdot m_{\tau}\bm{\nu} ds
    + [\delta\bm{r}\cdot m_{\nu}\bm{n}]
\end{equation}
where $\dot{\bm{v}}$ is the acceleration vector, $\rho_{0}$ is the density of the initial configuration, $\tau^{\alpha\beta}$ and $M_{0}^{\alpha\beta}$ are 2nd-Piola-Kirchhoff-like membrane stress and bending moment tensors, $\bm{f}$ and $\bm{t}$ are the body force and the traction, respectively, $m_{\tau}$ and $m_{\nu}$ are the normal and tangential components of the physical moment, and $\bm{\nu}$ is the normal vector to a parameterized curve cutting the surface at $\bm{r}$.

In this 2D simulation, the aforementioned Koiter model of non-Euclidean shells given in equation~(\ref{eq:incompatiblekoiter}) and (\ref{eq:incompatiblekoitermoduli}) below was used to calculate the membrane stress and bending moment tensors as well as the tangent tensors that are required in the consistent linearization of the weak form in equation~(\ref{eq:weakformfull}).  Here, we used silicone-based vinylpolysiloxane (VPS) as the sphere material, with $E$ = 1.3MPa and $\nu$ = 0.49 to impose the material incompressibility.
\begin{equation}
    \label{eq:incompatiblekoiter}
    \mathcal{U}=\frac{h}{2}\bar{\mathcal{A}}^{\alpha\beta\lambda\mu}\bar{\gamma}_{\alpha\beta}\bar{\gamma}_{\lambda\mu} +\frac{h^3}{24}\bar{\mathcal{A}}^{\alpha\beta\lambda\mu}\bar{\varrho}_{\alpha\beta}\bar{\varrho}_{\lambda\mu}
\end{equation}

\begin{equation}
	\label{eq:incompatiblekoitermoduli}
	\bar{\mathcal{A}}^{\alpha\beta\lambda\mu}=\frac{E}{2(1+\nu)}\left[\bar{a}^{\alpha\lambda}\bar{a}^{\beta\mu}+\bar{a}^{\alpha\mu}\bar{a}^{\beta\lambda}+\frac{2\nu}{1-\nu}\bar{a}^{\alpha\beta}\bar{a}^{\lambda\mu}\right]
\end{equation}
where $\bar{\mathcal{A}}$ is the tensor of elastic moduli of the non-Euclidean Koiter model.

The sphere was modeled as a bilayer shell in which the swelling ratio of the outer and the inner layers was different, {\em i.e.} differential swelling, but the value of natural curvature stimulus was calculated using the linear projection proposed by \cite{Pezzulla2017} as below in equation~(\ref{eq:linearprojectionl0}) and (\ref{eq:linearprojectionkappa}) such that the natural curvature stimulus was controlled via the difference between the swelling ratio of the outer and the inner layers and was imposed on the mid-surface of the sphere. 
\begin{equation}
    \label{eq:linearprojectionl0}
    \Lambda_{0} = \sqrt{\frac{m}{(1+m)}s_{\alpha}^{2} + \frac{1}{(1+m)}s_{\beta}^{2}}
\end{equation}

\begin{equation}
    \label{eq:linearprojectionkappa}
    \kappa = -\frac{1}{\Lambda_{0}^3}\frac{3}{h}\frac{m}{(1+m)^2}(s_{\alpha}^{2}-s_{\beta}^{2})
\end{equation}
where $m$ is the thickness ratio between the outer and inner layers, {\em i.e.} $m=\frac{h_{out}}{h_{in}}$, and $s_{\alpha}$ and $s_{\beta}$ are the swelling ratios of the outer layer and the inner layer, respectively (in the case without any swelling, $s_{\alpha}=s_{\beta}=1$). 

\begin{figure}[t] 
\begin{center}
\includegraphics[width=1\columnwidth]{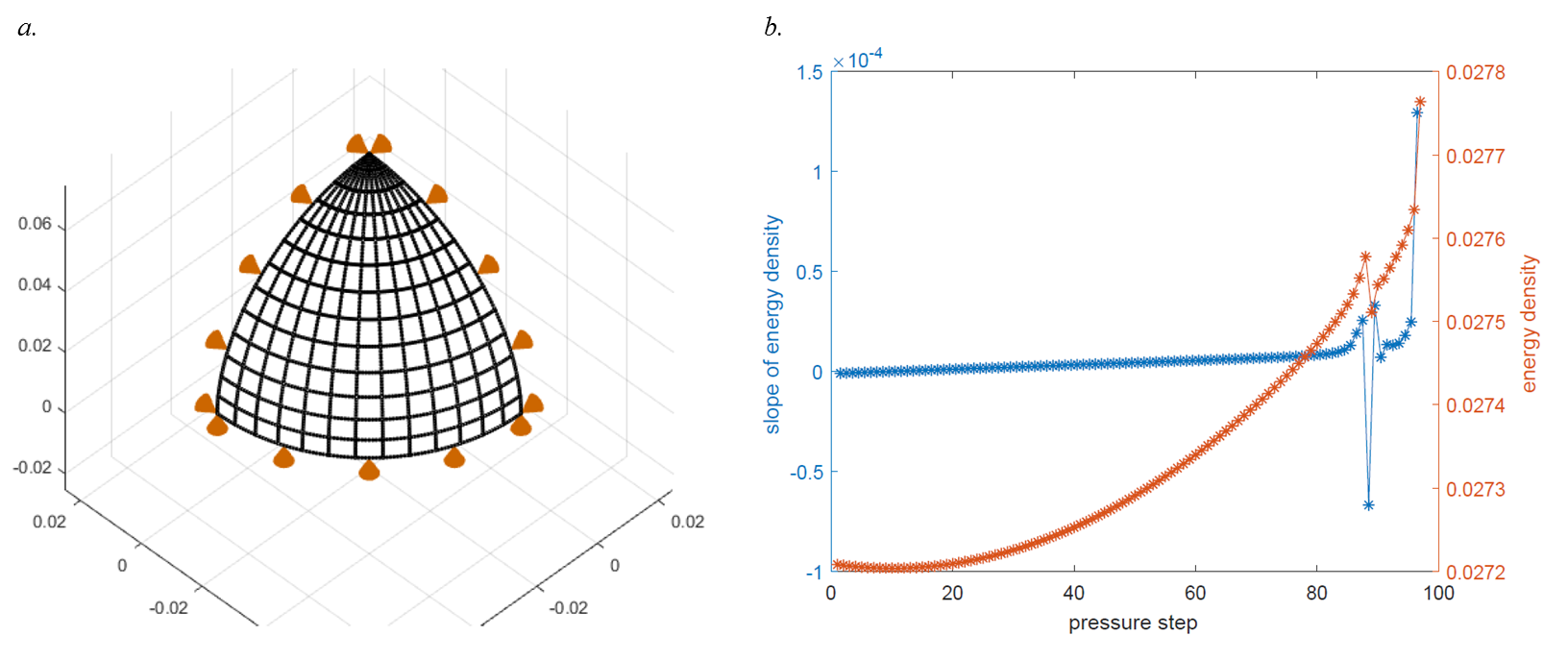}
\vspace{-6mm}
\caption{\label{fig:igasettup} $a.$ The 1/8 symmetric geometry model with brown arrows denoting its symmetry planes. $b.$ The critical points during multiple pressure steps due to the buckling.}
\end{center}
\end{figure}

Due to the symmetry shown in Fig.~\ref{fig:igasettup}a, 1/8 of the sphere was modeled in the IGA simulation, which was discretized by 13$\times$13 NURBS cubic elements.  The symmetry boundary conditions were applied on X=0, Y=0 and Z=0 in Fig.~\ref{fig:igasettup}a using the Lagrange multiplier method proposed by \cite{Duong2017} for the rotation constraints as described in equation~(\ref{eq:rotationconstraint}) and (\ref{eq:rotationconstraintg}) below 
\begin{equation}
    \label{eq:rotationconstraint}
    \Pi_{n}=\int_{\mathcal{L}_{0}} q(\bar{g}_{c}+\bar{g}_{s})dS
\end{equation}

\begin{equation}
    \label{eq:rotationconstraintg}
    \bar{g}_{c}=1-cos(\theta-\theta_{0}) 
    , \quad
    \bar{g}_{s}=sin(\theta-\theta_{0})
\end{equation}
where $q$ is the Lagrange multiplier, and $\theta$ and $\theta_{0}$ are the angles between the target geometry and the symmetry plane along the surface edge $\mathcal{L}$, for the current and initial configuration, respectively.  The pressure was applied as a live pressure using the unit normal vector of the area element of the current configuration.

In order to perform the nonlinear, quasistatic buckling analysis, the simulation was divided into two parts with ``multiple'' stimulus or load steps: first, the natural curvature stimulus was applied in the absence of pressure; and second, the pressure was imposed with the final value of natural curvature stimulus of the first part.  The increments of swelling ratio and pressure for the multiple steps were -0.05 and $p_{c}$/200, respectively.  Here, the increment of swelling ratio was added to the inner layer with $s_{\alpha}$ = 1 in the case of outward-pressure stimulus while it was added to the outer layer for inward-pressure stimulus with $s_{\beta}$ = 1.  To close the 2D simulation discussion, when the calculated value of the slope of strain energy of equation~(\ref{eq:incompatiblekoiter}) corresponding to the stimulus or load step has critical points, the value of stimulus or load for the first critical point was determined as the critical buckling stimulus or pressure $p_{\kappa}$ as shown in Fig.~\ref{fig:igasettup}b.


%

\end{document}